\documentclass[sigconf]{acmart} 
\AtBeginDocument{%
  }

\usepackage{booktabs}
\usepackage{multirow}
\usepackage{makecell}
\usepackage{graphicx}
\usepackage{siunitx}

\copyrightyear{2026}
\acmYear{2026}
\setcopyright{cc}
\setcctype{by}
\acmConference[IUI '26]{31st International Conference on Intelligent User Interfaces}{March 23--26, 2026}{Paphos, Cyprus}
\acmBooktitle{31st International Conference on Intelligent User Interfaces (IUI '26), March 23--26, 2026, Paphos, Cyprus}
\acmPrice{}
\acmDOI{10.1145/3742413.3789081}
\acmISBN{979-8-4007-1984-4/2026/03}

\begin{document}

\title{DesignBridge: Bridging Designer Expertise and User Preferences through AI-Enhanced Co-Design for Fashion}

\author{Yuheng Shao}
\authornote{Both authors contributed equally to this research.}
\orcid{0009-0008-6991-6427}
\affiliation{%
  \institution{School of Information Science and Technology, ShanghaiTech University}
  \city{Shanghai}
  \country{China}
}
\email{shaoyh2024@shanghaitech.edu.cn}

\author{Yuansong Xu}
\authornotemark[1]
\orcid{0009-0005-1630-6279}
\affiliation{%
  \institution{School of Information Science and Technology, ShanghaiTech University}
  \city{Shanghai}
  \country{China}}
\email{xuys2023@shanghaitech.edu.cn}

\author{Yifan Jin}
\orcid{0009-0007-3839-7612}
\affiliation{%
  \institution{School of Information Science and Technology, ShanghaiTech University}
  \city{Shanghai}
  \country{China}}
\email{jinyf2024@shanghaitech.edu.cn}

\author{Shuhao Zhang}
\orcid{0009-0008-1933-1869}
\affiliation{%
  \institution{School of Information Science and Technology, ShanghaiTech University}
  \city{Shanghai}
  \country{China}}
\email{zhangsh12024@shanghaitech.edu.cn}

\author{Wenxin Gu}
\orcid{0009-0005-3578-4479}
\affiliation{%
  \institution{School of Creativity and Art, ShanghaiTech University}
  \city{Shanghai}
  \country{China}}
\email{guwx2022@shanghaitech.edu.cn}

\author{Quan Li}
\authornote{Corresponding Author.}
\orcid{0000-0003-2249-0728}
\affiliation{%
  \institution{School of Information Science and Technology, ShanghaiTech University}
  \city{Shanghai}
  \country{China}
}
\email{liquan@shanghaitech.edu.cn}

\renewcommand{\shortauthors}{Yuheng Shao, Yuansong Xu et al.}

\begin{abstract}
  Effective collaboration between designers and users is important for fashion design, which can increase the user acceptance of fashion products and thereby create value. However, it remains an enduring challenge, as traditional designer-centric approaches restrict meaningful user participation, while user-driven methods demand design proficiency, often marginalizing professional creative judgment. Current co-design practices, including workshops and AI-assisted frameworks, struggle with low user engagement, inefficient preference collection, and difficulties in balancing user feedback with design considerations. To address these challenges, we conducted a formative study with designers and users experienced in co-design (N=7), identifying critical challenges for current collaboration between designers and users in the co-design process, and their requirements. Informed by these insights, we introduce \textit{DesignBridge}, a multi-platform AI-enhanced interactive system that bridges designer expertise and user preferences through three stages: (1) \textit{Initial Design Framing}, where designers define initial concepts. (2) \textit{Preference Expression Collection}, where users intuitively articulate preferences via interactive tools. (3) \textit{Preference-Integrated Design}, where designers use AI-assisted analytics to integrate feedback into cohesive designs. A user study demonstrates that \textit{DesignBridge} significantly enhances user preference collection and analysis, enabling designers to integrate diverse preferences with professional expertise.
\end{abstract}

\ccsdesc[500]{Human-centered computing~Interactive systems and tools}

\keywords{Co-Design, Collaboration, Preference Collection and Analysis}
\begin{teaserfigure}
 \centering 
 \includegraphics[width=\columnwidth]{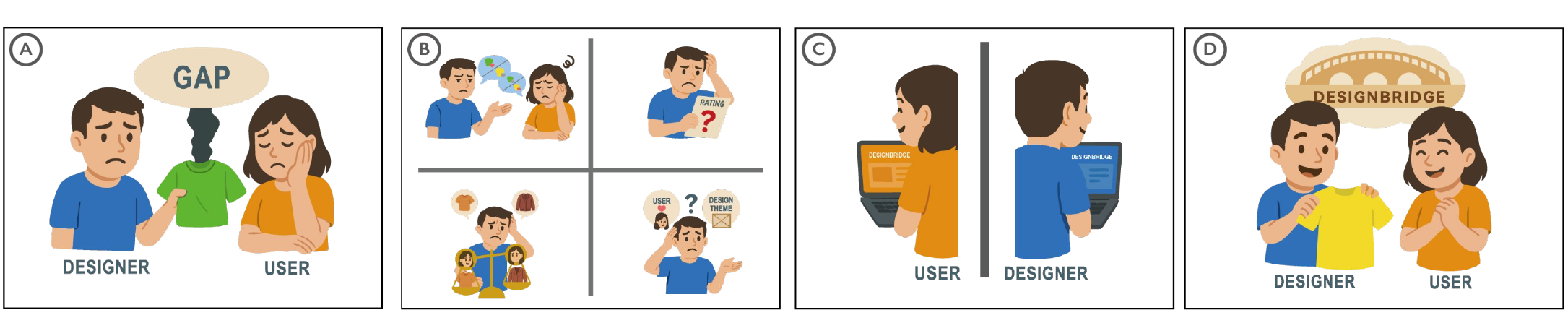}
 \caption{An example scenario illustrating how designers and users can effectively engage in a co-design process using \textit{DesignBridge}: (A) Designers and users face a gap in collaboration during co-design due to domain differences. (B) In fashion design, users often struggle to accurately and efficiently express their ideas, while designers must balance varying user preferences and navigate challenges such as trade-offs between user preferences and design concepts. (C) DesignBridge facilitates the process by assisting users in expressing their preferences and helping designers analyze user feedback for design integration. (D) This fosters collaboration between users and designers in the co-design process, enhancing both user experience and design outcomes.}
 \label{fig:teaser}
\end{teaserfigure}

\maketitle

\section{INTRODUCTION}
\par In design practice, professionals create products to meet user needs, while stakeholders act as both recipients and contributors through ongoing information exchange~\cite{vink2008defining}.
Designers typically elicit user requirements and preferences through analytical methods, surveys, and interviews to inform solution development~\cite{balloni2024social4fashion}. In fashion design, this process often follows an iterative \textit{Design-Display-Evaluation-Adjustment} cycle~\cite{hong2017interactive}, emphasizing coherent thematic concepts and user-centered scenarios. Following the establishment of an aesthetic direction and conceptual framework, designers explore multiple potential variations while considering user preferences. Unlike the production-preparation stage, where user involvement focuses on fit, sizing, and manufacturability~\cite{lee2015development,yang2024virtual,lee2025user}, this stage highlights the value of user collaboration during conceptual exploration and variation generation, thereby enhancing both relevance and creativity. Traditional approaches, however, remain predominantly design-centric, limiting direct user participation and restricting the articulation and integration of user needs within design frameworks. Although user-driven methodologies such as mass customization~\cite{da2001mass} and open design~\cite{buitenhuis2010open,bakirliouglu2019framing} empower users to steer design directions, they often require users to possess design expertise and may undermine the role of professional designers, risking diminished control over outputs and misalignment with predefined design concepts.


\par Existing co-design frameworks~\cite{steen2013co,zamenopoulos2018co} foster collaborative innovation. In fashion, research has explored co-design to augment designer creativity through human-AI collaboration~\cite{jeon2021fashionq,zhou2024understanding,jin2024understanding}. However, these studies focus primarily on designer-AI collaboration rather than direct user-designer interaction. Alternative modalities, such as workshops and collaborative creation, enable sustained interaction to align outputs with user needs and design concepts~\cite{zhang2022my, tullio2021empowering}. Nevertheless, these methods introduce operational complexities: extended timelines, communication inefficiencies, and difficulties reconciling non-expert input with professional workflows. Consequently, fashion design needs a more efficient, structured interaction paradigm to enhance user-designer collaboration.

\par To understand designer-user co-design challenges, we conducted a formative study with seven designers and experienced participants. Findings show that current processes struggle with effective preference collection and analysis. Low engagement and a lack of fine-grained methods result in incomplete preference capture, while the absence of mechanisms to balance user feedback with design considerations complicates the workflow. These insights inspire us to develop intuitive interactions that enhance user participation and enable designers to effectively fuse diverse feedback with their design intent.

\par Based on the insights from the formative study, we propose \textit{DesignBridge}, a multi-platform interactive system that supports the co-design process in the fashion design domain. The system has two interfaces: one for users and one for designers. The entire workflow is divided into three stages: \textbf{Initial Design Framing}, \textbf{Preference Expression Collection}, and \textbf{Preference-Integrated Design}. In \textbf{Stage One: Initial Design Framing}, designers input initial design considerations such as the design scene, clothing type, and principles. The system generates a scene image based on the input and filters relevant attributes from the design space to build the design image database. In \textbf{Stage Two: Preference Expression Collection}, users review provided garment designs on a virtual model in the given scene and a standalone garment image. Through brushing and commenting, users express their preferences (like or dislike) on specific parts or the overall design. Each user goes through multiple rounds of interaction to capture their overall preferences. In \textbf{Stage Three: Preference-Integrated Design}, designers examine the collected user preference information. Through a puzzle-like approach, they select attributes from each design space dimension to form a comprehensive design description based on the design space. For each attribute, designers navigate a hierarchical tree structure to examine the associated garments and user interactions. Through pruning, designers decide which attribute-related image data should be included in the generative model for fine-tuning. They can also review the model's predictions of individual user preferences for the current design, along with the model's interpretability results across different design space dimensions as a reference.

\par To evaluate \textit{DesignBridge}, we conduct a technical evaluation and a user study. The technical evaluation uses ratings to assess the effectiveness of the two fine-tuning processes in improving Design Space Capture and User Preference Integration, compared to the outputs generated by \textit{VisionRealistic v2 FluxDev}\footnote{\url{https://civitai.com/models/619656/vision-realistic}} and GPT-4o\footnote{\url{https://chatgpt.com}}. The user study further evaluates \textit{DesignBridge}'s effectiveness and usability from both the user and designer's perspective in the co-design process. The results indicate that the system could facilitate clear and efficient preference expression by users, as well as effective analysis of user feedback and integration of design considerations by designers during the design generation process. In summary, our contribution can be summarized as follows:

\begin{itemize}
    \item We conduct a formative study to identify the challenges and needs in the interactions between designer and users in the co-design process.
    \item We propose \textit{DesignBridge}, an interactive multi-interface system that supports the co-design process between designers and users.
    \item We conduct a technical evaluation and user study to collect the feedback from designers and users, evaluate the effectiveness and usability of \textit{DesignBridge}.
\end{itemize}

\section{BACKGROUND AND RELATED WORK}
\subsection{Interaction and Collaboration between Designer and Users}
\par In fashion design, the relationship between designers and users is inherently complex and dynamic. While designers bring their creative expertise to the development of fashion products, users contribute their individual needs, preferences, and expectations~\cite{siu2003users,haggman2015connections, zhou2024stylefactory, chen2024autospark}. The balance between these two perspectives is shaped by varying models of interaction and collaboration, including \textit{designer-initiative}, \textit{user-initiative}, and \textit{mixed-initiative} approaches.

\par In the \textit{designer-initiative} model, designers lead the process using their expertise and market analysis, with minimal user involvement. Traditional methods include surveys and data-driven analyses~\cite{furukawa2019visualisation,dubreuil2020traditional,an2020approaching}; for example, Furukawa et al.\cite{furukawa2019visualisation} used surveys to identify preferences, while An et al.\cite{an2020approaching} applied text mining for trend insights. Recent work integrates AI to further support designers~\cite{guo2023ai}, enhancing creativity~\cite{jeon2021fashionq} and enabling style fusion~\cite{wu2024stylewe}. Despite these advancements, excluding direct user participation risks misalignment with user expectations.

\par Conversely, the \textit{user-initiative} model places users to take the lead in the design process, with designers playing a supporting role in realizing user-defined ideas. A common example is crowdsourced fashion design, where users submit and vote on design proposals~\cite{di2014open, burton2012crowdsourcing, chung2021together}. While empowering, these approaches often suffer from a lack of standardization and extended design cycles, making them impractical for mainstream commercial adoption. To bridge this gap, some studies have explored ways to incorporate user input into commercial workflows—such as refining text prompts to generate personalized clothing images~\cite{liu2024mycloth} or adapting designs based on interactive user feedback~\cite{zhu2020interactive}. However, these techniques remain limited in scalability and integration, thus posing challenges for widespread application in commercial fashion design.

\par The mixed-initiative approach, grounded in research on human-computer interaction models~\cite{horvitz1999principles, cheng2025exploratory}, emphasizes the collaborative involvement of both designers and users in the co-design processes. In the context of fashion design, this paradigm facilitates joint participation, enabling both parties to contribute their perspectives and expertise~\cite{zhang2022my, peterson2016co, ulrich2003consumer, tullio2021empowering, sharma2020implementation}. Traditional implementations of mixed-initiative design include co-creation workshops~\cite{tullio2021empowering} and collaborative prototyping~\cite{konola2014co}, often supported by digital platforms such as \textit{Figma}~\cite{figma} and \textit{Uizard}~\cite{uizard}, which integrate AI technologies to assist in design ideation and iteration. While these tools promote creativity and shared ownership, they typically require multiple design cycles, leading to extended timelines and reduced efficiency. To align design outputs with user needs, some studies have introduced the use of design probes~\cite{wallace2013making}, which leverage users' everyday experiences and inspirations as contextual feedback. However, this approach often encounters difficulties in interpretation due to the gap between user expressions and designer comprehension. More recent work has explored value co-creation in co-design~\cite{zhang2021co}, incorporating technologies like Virtual Reality (VR) to enable immersive user engagement and real-time collaboration with designers~\cite{yang2024virtual,sarakatsanos2024vr}. However, challenges persist in effectively eliciting user needs and bridging communication gaps in differences of professionalism and design literacy.

\par Our work adopts a mixed-initiative paradigm within the context of fashion design, aiming to enhance designer-user collaboration throughout the co-design process. We enable users to articulate their needs without the constraints of professional terminology, while assisting designers in accurately interpreting and incorporating those needs. Furthermore, we refine the generative model based on user preferences, integrating these insights into the design outcomes to improve the overall effectiveness and personalization of the co-design experience.

\subsection{Preference Elicitation and Synthesis}
\par Understanding and incorporating user needs is fundamental to achieving high-quality design outcomes and ensuring user acceptance~\cite{courage2005understanding,veryzer2005impact}. A critical component of this process is effective preference handling, which ensures that user needs are accurately captured and appropriately reflected in the final design~\cite{patel2016preference,miettinen2015new}. This process typically involves two key steps: \textit{preference elicitation} and \textit{preference synthesis}.

\par Preference elicitation refers to the process of gathering information about users' preferences and is generally categorized into two approaches: \textit{explicit elicitation} and \textit{implicit inference}~\cite{salamo2012generating, wang2025prefer2sd}. Explicit elicitation requires users to actively express their preference, for example through ratings, rankings, or questionnaires~\cite{jameson2004more, o2001polylens}. AI Tailoring~\cite{li2024ai}, for instance, employs triplet preference ranking to extract user preferences, while Zeng et al.~\cite{zeng2024intenttuner} introduced \textit{IntentTuner}, a framework that refines design preferences via multimodal inputs and interactive feedback. In contrast, implicit inference involves analyzing user behavior data—such as interaction patterns, historical activity, or feedback—to uncover latent preferences~\cite{crossen2002flytrap, lieberman1998let}. E-commerce platforms like \textit{Amazon}\footnote{\url{https://www.amazon.sg}} exemplify this approach by utilizing recommendation systems that infer preferences from click behavior, dwell time, and purchase history to personalize content delivery.

\par Once preferences have been elicited, the next step is preference synthesis, which involves modeling, integrating, and analyzing the gathered information and feeding it back into the design decision-making process~\cite{salamo2012generating, tran2024overview}. This step often requires resolving various complexities and potential conflicts in user preferences~\cite{hasan2006conflict,nurgaliyev2017improved}. Recent studies have explored more sophisticated techniques for preference synthesis~\cite{salamo2012generating, shin2022chatbots, becattini2024interactive}. For instance, Becattini et al.~\cite{becattini2024interactive} integrated explicit feedback with implicit inference using reinforcement learning agents in a clothing recommendation system. Salamó et al.~\cite{salamo2012generating} provided a comprehensive overview of consensus-building strategies to resolve preference conflicts within and across user groups. Additionally, chatbots have been explored as tools to support asynchronous co-design by facilitating communication and consensus-building among distributed stakeholders~\cite{shin2022chatbots, chandrasegaran2025synthetic}.

\par Building on semantic classification methods, our work proposes a multimodal, recommendation-driven mechanism for collecting user preferences. This approach employs utility area and machine learning algorithms to model preferences at both the individual user and design element levels. To improve transparency and user trust, we also integrate explainability techniques that enhance the interpretability of the entire process.

\subsection{Generative Tools in Fashion Design}
\par The rapid advancement of deep learning and large-scale pre-trained models has led to the emergence of generative interactive tools and interfaces, which are transforming contemporary design workflows~\cite{kahng2018gan,wang2023large,jeon2021fashionq}. Traditional digital design often involves iterative transitions between conceptualization, prototyping, and refinement stages, requiring significant manual effort and cognitive load. In contrast, generative techniques—powered by diffusion models~\cite{croitoru2023diffusion,yang2023diffusion} and large language models (LLMs)~\cite{chang2024survey}—enable rapid synthesis of sketches or prototypes from minimal prompts~\cite{adamkiewicz2025promptmap}, streamlining design exploration and diversifying creative possibilities.

\par A growing body of research has explored generative interfaces in the context of garment design, with a primary emphasis on supporting designer creativity. For example, \textit{StyleMe}~\cite{wu2023styleme} produces designer-specific stylistic sketches to inspire fashion ideation, while \textit{StyleWe}~\cite{wu2024stylewe} facilitates the integration of diverse design styles into new creations to encourage collaborative inspiration. Jiang et al.~\cite{jiang2024haigen} proposed a human-AI collaborative system aimed at improving design efficiency. Despite these advancements, most existing tools are tailored to enhance professional design capabilities, with limited emphasis on understanding and incorporating target users' preferences. Although some systems reduce skill barriers—such as \textit{MYCloth}~\cite{liu2024mycloth}, which leverages Stable Diffusion to support personalized T-shirt customization—these solutions often lack the depth of designer expertise, leading to inconsistencies in generative quality. Furthermore, they fail to holistically integrate interface design, algorithmic transparency, and user experience optimization across a diverse target user base~\cite{guo2023ai}.

\par To bridge these gaps, we propose a three-stage interactive framework that seamlessly integrates generative models with both designer expertise and user feedback. Grounded in a comprehensive, multi-dimensional design space that encapsulates key attributes of fashion design, our approach supports iterative co-creation. Designers initiate and refine the generative process, while users contribute real-time feedback to guide outcomes. Through this adaptive interaction loop, the system continuously evolves to align generative outputs with both aesthetic intent and user expectations, achieving a more balanced integration of creativity, usability, and personalization.

\section{FORMATIVE STUDY}
\subsection{Methodology}
\par Our primary objective is to refine the interaction paradigm between designers and users to facilitate the collaborative design process. To inform this goal, we conducted a formative study with institutional IRB approval to investigate existing interaction paradigms in design practice and identify challenges encountered by both stakeholders.

\par We recruited seven qualified participants (3 female, 4 male) drawn from a local company or as users engaged in the company's design initiatives. All possessed prior co-design experience. The participant group comprised three professional designers (designated P1–P3) employed by a commercial clothing brand, each with over three years of professional design experience, and four users who had previously collaborated with the same brand in fashion co-design projects.


\par We conducted 40-minute semi-structured interviews with each participant via online video conferencing. For designers, discussions centered on their collaborative workflow with users, strategies for eliciting user needs, and balancing user input with creative vision. For users, interviews focused on their prior co-design experiences, including perceived engagement, satisfaction, and the clarity of need articulation during collaboration. Interview protocols are outlined in \autoref{interview-questions}, and all sessions were video-recorded. We employed thematic analysis~\cite{braun2006using,braun2012thematic} to interpret the data: transcripts were generated from audio recordings, reviewed by the first author, and iteratively analyzed by two authors to identify, categorize, and synthesize emergent themes and codes.


\begin{table}[h]
\centering
\small
\setlength{\tabcolsep}{4pt}
\renewcommand{\arraystretch}{1.05}
\caption{Demographic information of participants in the formative study.}
\label{tab:formative-participants}
\begin{tabular}{
  m{0.9cm}
  m{1.5cm}
  m{1.4cm}
  m{1.6cm}
  m{1.7cm}
}
\toprule
\textbf{ID} &
\textbf{Role} &
\textbf{Gender} &
\textbf{Age} &
\textbf{\shortstack{Experience /\\ Familiarity}} \\
\midrule
P1 & Designer & Female & 33 & 5 years \\
P2 & Designer & Male & 36 & 6 years \\
P3 & Designer & Male & 32 & 3 years \\
P4 & User & Female & 25 & High \\
P5 & User & Male & 27 & Moderate \\
P6 & User & Female & 24 & High \\
P7 & User & Male & 26 & Moderate \\
\bottomrule
\end{tabular}
\end{table}

\subsection{Findings}






\par Based on the results, we summarized the current co-design workflow as a three-stage process. In \textbf{Stage One: Initial Design Framing}, designers establish a preliminary framework that articulates the aesthetic direction, stylistic elements, and thematic concepts, often through tools such as moodboards and concept sketches. In \textbf{Stage Two: Preference Expression Collection}, designers generate alternative variations using methods such as generative models or prototyping tools—and collect user feedback through mechanisms such as annotated selections in Figma or comparative evaluations of moodboards. Finally, in \textbf{Stage Three: Preference-Integrated Design}, designers synthesize user feedback with the established framework to iteratively refine the design, adjusting features such as color palettes or textures to align with user expectations while preserving the integrity of the original creative vision. During this structured process, several challenges persist:

\subsubsection{C1. Ineffective Preference Collection Stemming from Low Engagement and Suboptimal Interaction}
\par A primary challenge in co-design processes is the inadequate capture of user preferences, often exacerbated by limited engagement and contextual immersion. Designers first recognized that ``\textit{Traditional methods like surveys just passively gather user info.}'' (P2). As a result, the current approach typically involves users in design workshops and other interactive sessions. However, existing methodologies frequently fail to situate users within the thematic or situational context of the design, hindering their ability to provide nuanced feedback. As P3 noted: ``\textit{They told me we're designing winter sportswear, but outside it's summer. How can I possibly envision myself wearing an outdoor jacket, braving the cold wind in the snow?}'' ``\textit{......Although I have some experience in both design participation and personal purchases, I find it sometimes difficult to fully put myself into others' shoes, making my feedback naturally inaccurate.}''
\par Further inefficiencies arise from mismatched knowledge backgrounds and insufficient guidance during designer-user interactions. Users often articulate ideas intuitively rather than technically, while designers face challenges translating these inputs into actionable insights. As a designer, P2 explained, ``\textit{Users usually describe their thoughts in a more intuitive, casual way rather than using technical terms, such as: `I don’t like this pattern; I think it should feel more mature and sophisticated.' Without someone effectively bridging the gap and turning these intuitive ideas into concrete design considerations, communication gets stuck.}'' This underscores the need for mediating frameworks to bridge communication gaps and enhance mutual understanding.

\subsubsection{C2. Coarse-Grained Preference Capture via Qualitative Methods}
\par Both designers and users expressed concerns about the coarse-grained and insufficient collection of individual preferences in existing methods, which rely primarily on qualitative methods such as interviews and discussions. These approaches inadequately resolve the multidimensionality of preferences. For example, P7 remarked, ``\textit{I think the way we interact right now has its limits because how much of our preferences get captured really depends on how the discussion goes. If they ask me, `Do you like this style?' I might respond, `No, it looks old-fashioned.' But in reality, I am not entirely rejecting it—I actually like certain elements, such as the fitted cuffs and the central pattern. However, when combined, they feel unappealing.}'' Designers also pointed out that user preferences are rarely binary but often complex and multidimensional. ``\textit{If we assume that users dislike shirts simply because they gave negative feedback on the designs we presented, this could be misleading. The problem might not be with the shirts themselves, but with the specific design we showed. For example, if we only showed high-collared business shirts, people might reject them because they feel too formal, but they could be totally fine with casual round-neck or knit-collar shirts.}'' (P1). These findings advocate for finer-grained design-space definitions and precision-driven preference elicitation.

\subsubsection{C3. Demand for Multidimensional Feedback Integration in Iterative Design}
\par During the analysis process, users may not always be consistent in their preferences, which can be diverse and even contradictory. If only the common majority-rule approach is used, it may not be the most rational method. Preferences are often multi-dimensional, focusing on specific aspects, such as collar style, rather than the overall design. As a result, users might express that they ``\textit{like the overall design but are dissatisfied with certain parts}'' (P3), making an overall vote insufficiently accurate. Moreover, the designer expressed the need to understand users' feedback on design iterations. As pointed out by P3, ``\textit{We need to ensure that diverse user opinions are considered rather than simply following the majority. Ideally, I want to be able to see how much users actually like my design and understand how changing specific details, like the cuffs or materials, would affect their preferences. This would really help improve the design process, but right now, the co-design approach doesn’t capture these insights well.}''


\subsubsection{C4. Balancing Design Intent with User Preferences}
\par In the design decision-making process, beyond considering user opinions, another crucial aspect is integrating the design theme and context. It is essential to effectively combine both elements to ensure that user preferences are thoroughly considered within the initial design framework, ultimately leading to a well-informed design outcome. However, challenges remain in this process, as designers lack a unified guideline on whether and to what extent all user preferences should be incorporated. As P2 stated, ``\textit{In the current process, we typically analyze the collected user feedback and then develop design solutions based on it. I acknowledge that in this process, how we reference and to what extent we incorporate user feedback often relies on subjective judgment, which may lead to incompleteness or bias.}'' Given these considerations, a systematic approach is needed to comprehensively integrate user needs with design considerations, ultimately enhancing the design decision-making outcome.

\subsection{Design Goals}
\par Informed by these insights, we formulated four design goals to underpin the development of our framework. Our methodology centers on optimizing the collaborative interaction between designers and users in the identified three-stage co-design workflow, from the construction of an initial design framework, collecting users' preferences, and analysis of user preferences alongside iterative design refinement.

\par \textbf{DG1. Enhance User Engagement Through Intuitive Design-User Interaction.} To resolve \textbf{C1}, the system should present preliminary design considerations clearly and intuitively to enhance user engagement and effective preference collection. This requires fostering intuitive interactions that align with users' everyday decision-making processes (e.g., garment selection and purchasing). For instance, enabling users to denote preferences for both holistic designs and specific components (e.g., collars, patterns) through familiar, gesture-driven interfaces could bridge communication gaps while maintaining contextual relevance.


\par \textbf{DG2. Support Accurate and Efficient User Preference Capture.} To resolve \textbf{C2}, the system must establish a well-defined design space to ensure precise interpretation of user inputs. This involves integrating quantitative, fine-grained preference-capture mechanisms during collaborative sessions, enabling designers to systematically explore multidimensional preferences (e.g., stylistic elements, functional attributes) through structured, data-informed workflows.

\par \textbf{DG3. Provide Comprehensive Consideration of User Feedback Analysis.} Addressing \textbf{C3}, the system should synthesize diverse—and potentially conflicting—user feedback via consensus-building algorithms, presenting results through intuitive, interactive visualizations. Additionally, it must provide dynamic feedback on design iterations, such as predictive acceptance metrics and consensus levels, to support evidence-driven adjustments.

\par \textbf{DG4. Balance Design Consideration with User Feedback.} To assist in balancing design considerations and user feedback during the co-design process (\textbf{C4}), the system should embed core design themes and contextual scenarios into the interaction workflow. When generating designs, designers should be allowed to make an informed decision on which user feedback should be incorporated into the generative model, fine-tuning the results to ensure the design balances design considerations with user feedback.

\subsection{Garment Design Space}
\label{sec:34}
\par
\begin{figure}[h]
  \centering
  \includegraphics[width=\linewidth]{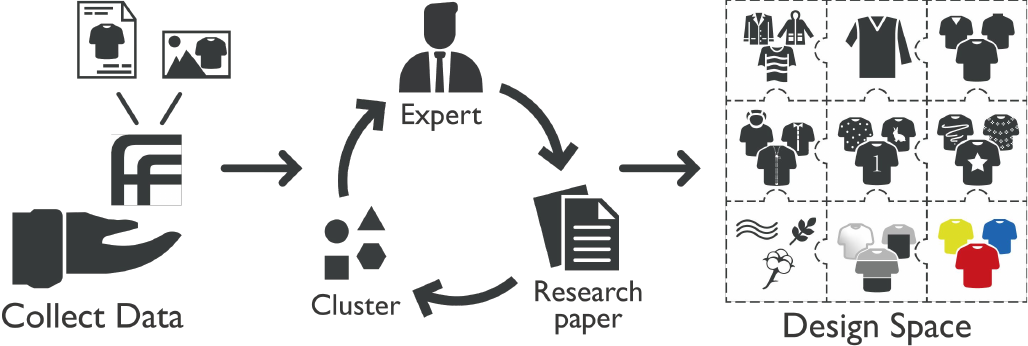}
  \caption{The entire process of the identification of garment design space, including collecting design image with corresponding description, conduct clustering analysis incorporating expert knowledge and research, and identify the definition of garment design space.}
  \label{fig:design space}
\end{figure}

 To address design goal \textbf{DG2}, we established a structured design space for fashion, which serves as a foundation for both collecting user preferences and supporting fashion generation tasks. \textbf{P1}–\textbf{P3} continued their collaboration with us throughout this process. The complete workflow is illustrated in \autoref{fig:design space}. We began by curating a dataset of $3,294$ fashion samples from \textit{Farfetch}\footnote{\url{https://www.farfetch.com}}, a leading fashion e-commerce platform. Each sample included a high-quality image of an upper-body garment against a white background, accompanied by a textual description—providing both visual and semantic information.

\par To prepare the textual data, we first applied regular expressions to remove redundant prepositions. We then performed clustering on $16,472$ text samples using the \textit{MinHashLSH} method~\cite{broder1997resemblance,gionis1999similarity}, yielding $123$ clusters. These clusters revealed patterns in the real-world data and served as a starting point for expert review. During the collaborative evaluation, \textbf{P2} noted, ''\textit{The clustering results are meaningful, but there are still issues. For example, items are grouped by color keywords, which leads to yellow collars and yellow sleeves being clustered together, even though they represent distinct design elements.}'' The experts also referred to prior fashion design research~\cite{ma2017towards, liu2016deepfashion, jeon2021fancy} to support their analysis. As \textbf{P3} remarked, ``\textit{The literature helps me recall more precise terms for describing observed design elements. It also confirms my intuition that abstract attributes like `casual' or `formal' should be excluded from the design space, as they can typically be inferred from combinations of more concrete features.}'' While large-scale datasets such as \textit{DeepFashion}~\cite{liu2016deepfashion} and \textit{FANCY}~\cite{jeon2021fancy} define hundreds of fine-grained attributes, our experts emphasized that adopting such granularity in interactive co-design could impose excessive cognitive burden on users and complicate preference expression. Based on this consideration, a moderately granular design space was chosen to balance descriptive precision with usability, while finer-grained control was reserved for designers in later stages of the workflow.

\par Through this iterative expert-in-the-loop process, we finalized a structured design space comprising nine key dimensions: \textit{Type}, \textit{Sleeve Length}, \textit{Collar Shape}, \textit{Wearing Style}, \textit{Pattern Style}, \textit{Pattern Arrangement}, \textit{Material}, \textit{Color Category}, and \textit{Specific Colors}. Each dimension includes multiple attribute values. The design space is detailed in \autoref{sec:design space}.

\section{DESIGNBRIDGE}

\subsection{Overview}
\par
\begin{figure*}[h]
    \centering
    \includegraphics[width=\textwidth]{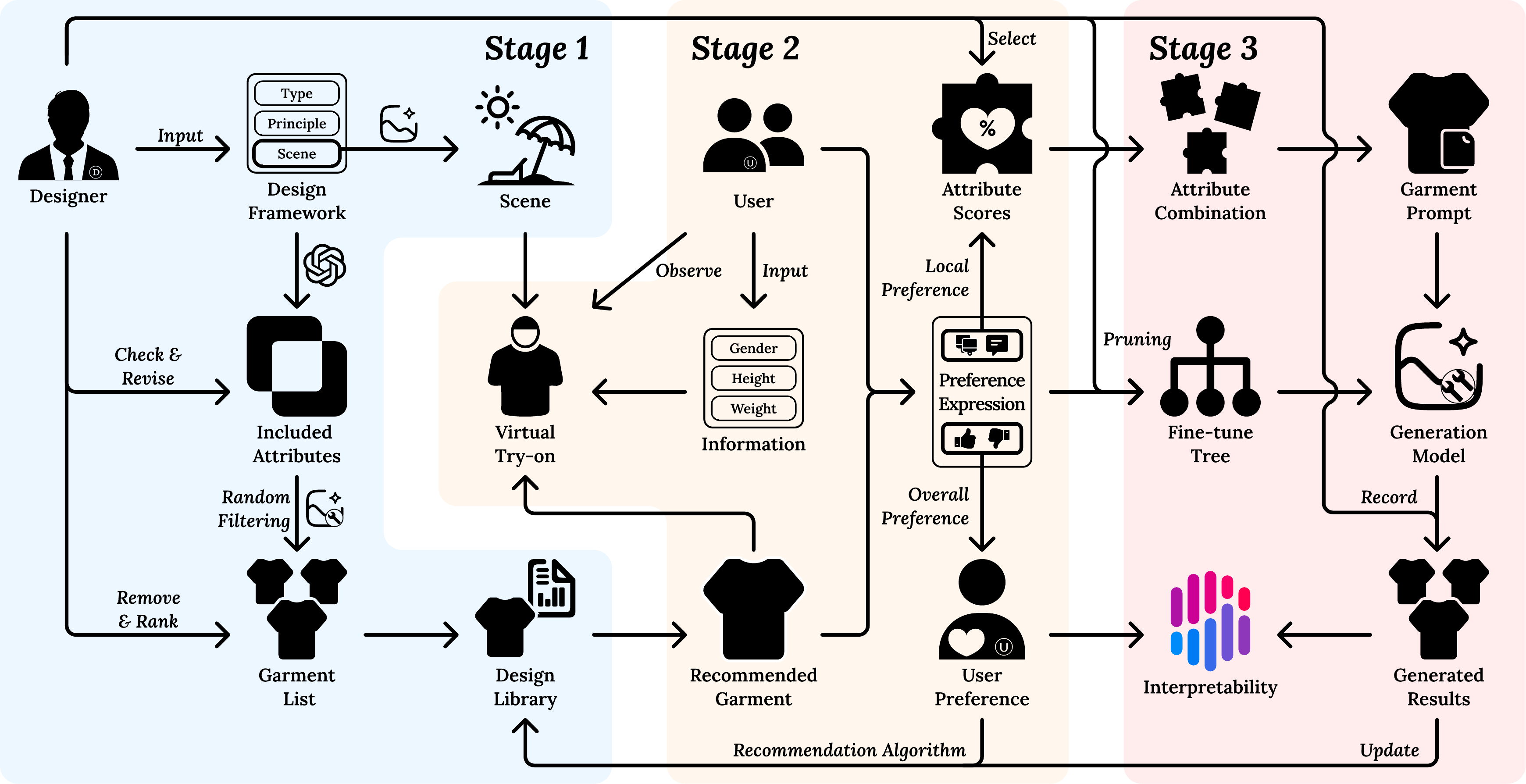}
    \caption{The workflow of \textit{DesignBridge} is structured into three stages. In Stage One, designers provide initial inputs to identify a suitable design framework and generate corresponding images. In Stage Two, users evaluate the designs by viewing them both on a virtual model within a contextual scene and against a neutral background, expressing preferences at both the local and global levels. In Stage Three, designers analyze the collected user feedback, synthesize attributes across different dimensions of the design space, and iteratively refine the designs based on predicted user responses.}
    \label{fig:workflow}

\end{figure*}

 As illustrated in \autoref{fig:workflow}, \textit{DesignBridge} is a multi-interface system that facilitates co-design through three sequential stages: \textbf{Initial Design Framing}, \textbf{Preference Expression Collection}, and \textbf{Preference-Integrated Design}. The system includes dedicated interfaces for both designers and users to support a collaborative and iterative design process.

\par In Stage One: \textbf{Initial Design Framing}, designers operate within the \textit{Framing Panel View} to input initial design parameters such as the target scenario, clothing type, and relevant design principles. Based on these inputs, the system constructs an initial design framework that filters out incompatible attributes from the broader design space. Prior studies show that designers often employ generative models to create initial design proposals~\cite{jiao2021role,chen2019pog} to collect board user feedback. Building on this practice, the system leverages this framework to generate a representative scene image and a set of preliminary design images, forming a fashion design database. This transformation of abstract design intent into visual and contextualized outputs fosters user engagement from the outset of the co-design process \textbf{(DG1)}.

\par In Stage Two: \textbf{Preference Expression Collection}, users interact with the system through the \textit{Information View}, where design images are displayed via virtual try-on within the scene context. Simultaneously, the \textit{Interaction View} presents garment images on a neutral background for focused evaluation. During each interaction round, users express their preferences by indicating overall likes or dislikes and using a brush tool to highlight specific areas, supplemented by textual comments for clarification. After multiple rounds, \textit{DesignBridge} aggregates this feedback to construct a detailed profile of user preferences. This stage combines intuitive interaction with fine-grained input mechanisms, supporting precise capture and interpretation of user intent \textbf{(DG1, DG2)}.

\par In Stage Three: \textbf{Preference-Integrated Design}, designers review the consolidated user feedback, including ranked attribute preferences within each design dimension based on consensus analysis. This helps guide the selection of key attributes for constructing a final design by referring to the consensus of user preferences \textbf{(DG3)}. Designers can also inspect the design images associated with each attribute and examine individual user feedback to determine how these preferences should influence model-generated outputs, balancing user preferences with design consideration \textbf{(DG4)}. Furthermore, by analyzing the predicted user preferences and understanding how each design dimension contributes to the system's predictions, designers can iteratively refine their outputs. This enables continuous improvement and adaptation throughout the design process.

\subsection{Initial Design Framing}

\begin{figure*}[h]
    \centering
    \includegraphics[width=\textwidth]{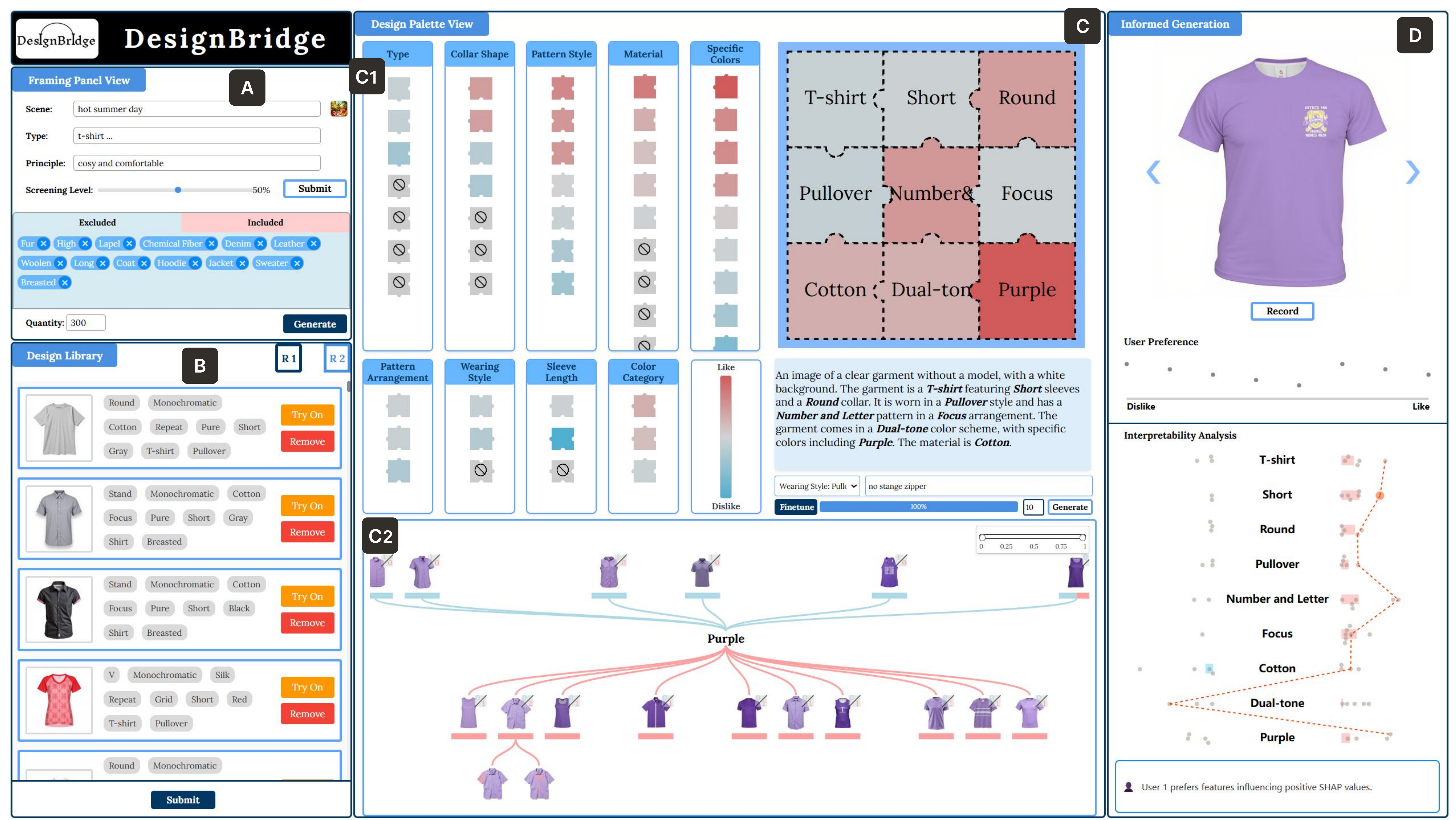}
    \caption{The Designer Interface of \textit{DesignBridge} consists of four views: (A) The \textit{Framing Panel View} receives the designer’s initial input to construct a preliminary design framework. (B) The \textit{Design Library View} builds a design image database based on the design framework. (C) The \textit{Design Palette View} integrates user feedback to assist designers in constructing the design. (D) The \textit{Informed Generation View} presents the generated design images and the predicted user preference feedback of the current design.}
    \label{fig:designer interface}
\end{figure*}

\subsubsection{Design context framing}
\par In the designer interface, the \textit{Framing Panel View} (\autoref{fig:designer interface}-A) provides an input section where designers specify key parameters—\textit{Type}, \textit{Scene}, and \textit{Principles}—to construct the initial design framework. The \textit{Type} field defines the garment category, selected from seven predefined types listed in \autoref{tab:design space}. The \textit{Scene} field captures the intended usage context, such as ``formal business'' or ``sport outdoor'', while the \textit{Principle} field allows designers to articulate high-level intentions, such as ``functional, comfortable, and durable'' or ``minimalist style''.

\par Upon submission via the \raisebox{-0.4ex}{\includegraphics[height=2.3ex]{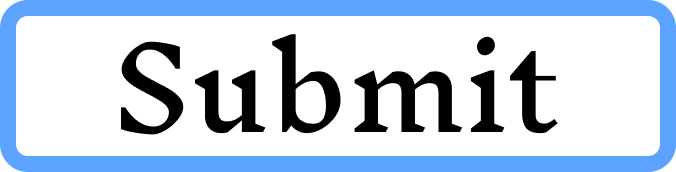}}, the system generates a reference scene image based on the scene description. A thumbnail of the image is displayed to the right of the scene input, and clicking it reveals the full-size version. The system also analyzes the combined input to filter incompatible attributes across multiple design dimensions. For example, if ``\textit{formal attire}'' is specified in the design principles, the system may exclude pattern attributes like ``\textit{dot}'' or ``\textit{number and letter}'' as misaligned with the intended style.

\par Designers can further refine this filtering process using the \textit{filter slider}, which adjusts the strictness of attribute selection—higher values apply more stringent criteria. The results display both included and excluded attributes as selectable labels beneath the input section, enabling designers to review and toggle the inclusion status of any attribute by simply clicking its corresponding label.

\subsubsection{Design image database construction}
\par After finalizing the filtered attributes, designers can specify the desired number of design outputs using the quantity input field and initiate image generation by clicking the \raisebox{-0.4ex}{\includegraphics[height=2.3ex]{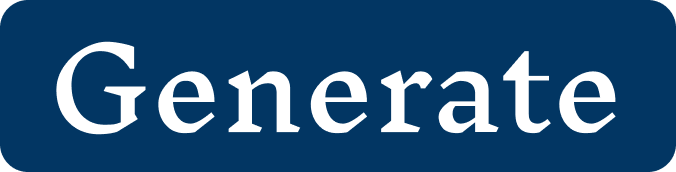}}. The system then produces garment images that align with the defined design context. These generated designs are displayed in the \textit{Design Ligrary} (\autoref{fig:designer interface}-B), where designers can review them alongside their corresponding attributes across multiple design space dimensions.

\par Designers can preview how each garment appears on a virtual model within the specified scene using the \raisebox{-0.4ex}{\includegraphics[height=2.3ex]{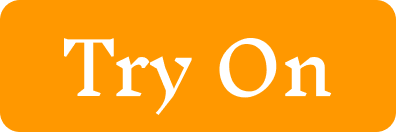}}. The interface also supports interactive refinements—designers may remove specific images using the \raisebox{-0.4ex}{\includegraphics[height=2.3ex]{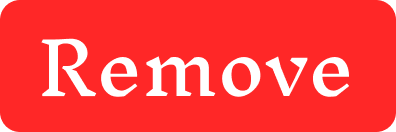}} or reorder them through drag-and-drop operations to better organize their selections. Once the review is complete, clicking the \raisebox{-0.4ex}{\includegraphics[height=2.3ex]{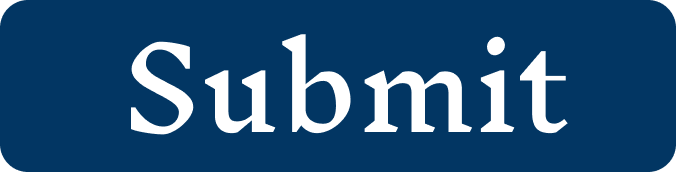}} in the lower right corner submits the finalized design images to the backend, where they are stored in the database for subsequent use within the user interface.

\begin{figure*}[h]
    \centering
    \includegraphics[width=\textwidth]{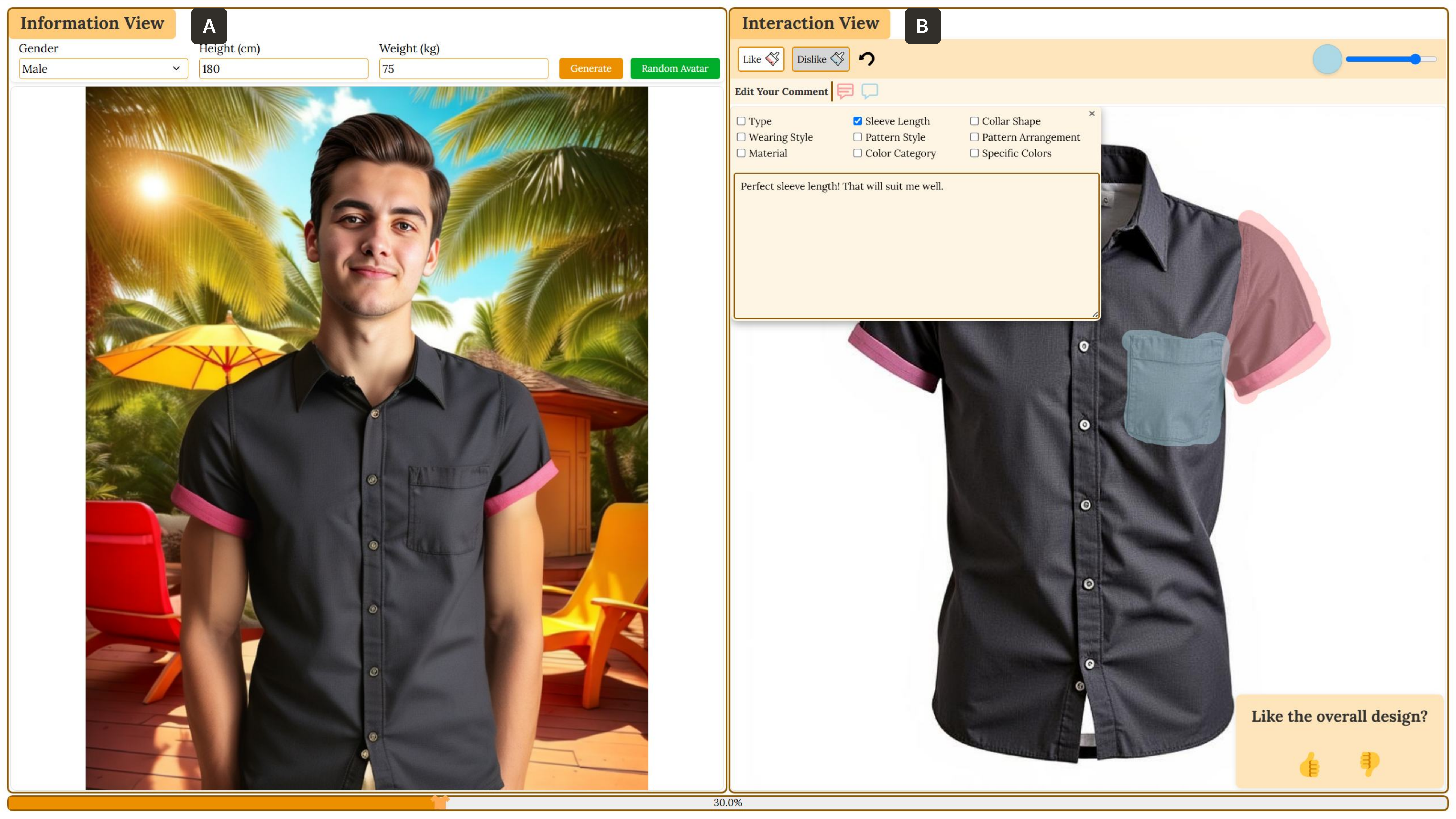}
    \caption{The User Interface of \textit{DesignBridge} consists of two views: (A) The \textit{Information View} receives the user’s information to construct a virtual model for design try-on in the given scene. (B) The \textit{Interaction View} enables users to express preferences for each design.}
    \label{fig:user interface}
\end{figure*}

\subsection{Preference Expression Collection}
\subsubsection{Information collection and user preference expression}
\par Within the  \textit{Information View} (\autoref{fig:user interface}-A) in the user interface, users can input anthropometric information such as height, weight, and gender in the \textit{Information View} or randomly generate a model by clicking \raisebox{-0.4ex}{\includegraphics[height=3.1ex]{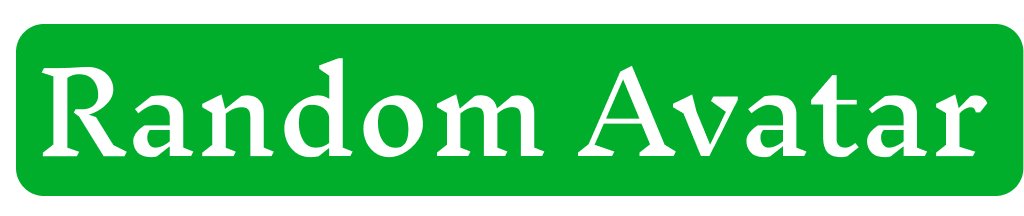}}. This information is used to generate a personalized virtual model for garment display. Upon clicking the 
\raisebox{-0.4ex}{\includegraphics[height=2.3ex]{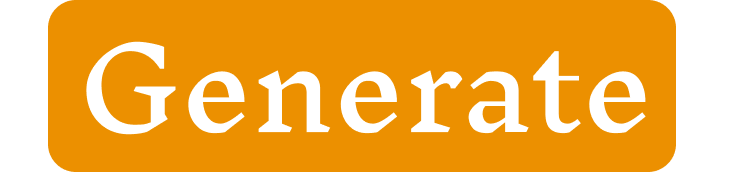}}, the system displays the current garment design on the virtual model situated within the specified scene context.

\par Below the input section, the \textit{Information View} (\autoref{fig:user interface}-B) presents a try-on preview, allowing users to see how the garment appears in context. On the right side, the \textit{Interaction View} displays a standalone garment image against a blank background. Users can provide detailed feedback by using the \raisebox{-0.4ex}{\includegraphics[height=2.3ex]{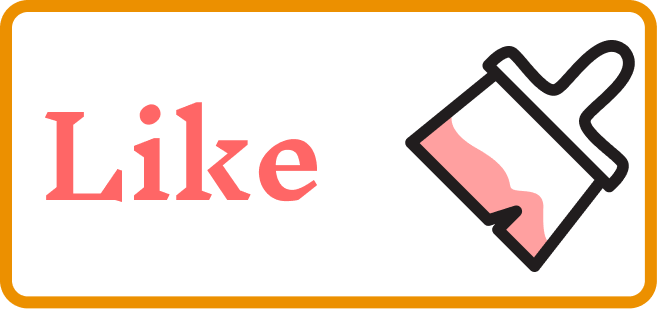}} or \raisebox{-0.4ex}{\includegraphics[height=2.3ex]{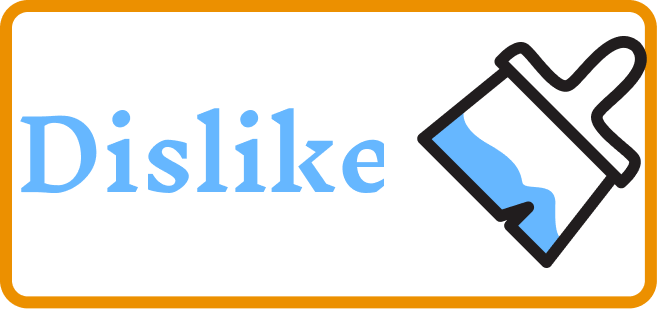}} to highlight specific garment areas. For instance, when using the \raisebox{-0.4ex}{\includegraphics[height=2.3ex]{figures/like_button.png}}, users can drag over the garment image to apply a colored overlay, indicating preferred regions. Upon completion of a brush action, a \raisebox{-0.4ex}{\includegraphics[height=2.3ex]{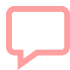}} or \raisebox{-0.4ex}{\includegraphics[height=2.3ex]{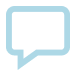}} appears in the upper-left corner of the view. Clicking this icon opens an information panel displaying a checklist of nine design dimensions.

\par Leveraging a text-image comparison model and pixel-level analysis, the system suggests the most relevant design attributes associated with the brushed area, which users can confirm or modify. For example, if a user brushes the sleeve, the model identifies its location and semantic content, analyzes visual properties such as size and shape, and suggests that the interaction likely refers to the ``sleeve length'' attribute. Users may also supplement their feedback with free-text comments. Additional controls include an \raisebox{-0.4ex}{\includegraphics[height=2.3ex]{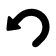}} to remove brush actions and a slider to adjust brush size.

\par In addition to localized feedback, users can express an overall opinion by selecting the \raisebox{-0.4ex}{\includegraphics[height=2.3ex]{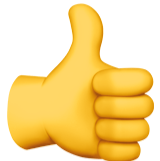}} or \raisebox{-0.4ex}{\includegraphics[height=2.3ex]{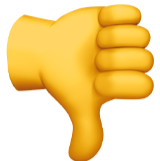}} button located in the bottom-right corner. Upon submission, the system presents the next garment design for evaluation. Through multiple rounds of interaction, the system iteratively learns and refines a comprehensive, multidimensional representation of the user's preferences within the current design context.

\subsection{Preference-Integrated Design}
\subsubsection{Constructing design entities by tuning based on user feedback}
\par In the designer interface, designers can explore attributes across various design space dimensions through the \textit{Design Palette View} (\autoref{fig:designer interface}-C). Each dimension presents attributes as vertically aligned \raisebox{-0.4ex}{\includegraphics[height=2.3ex]{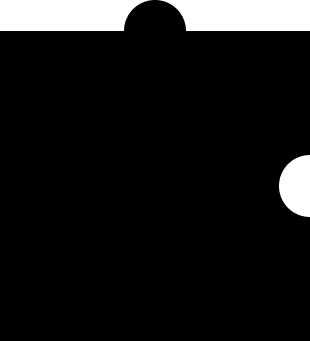}} of uniform shape, organized by consensus scores derived from the system. The color gradients of these pieces indicate the distribution of these scores. By hovering over a puzzle piece, designers can view both the attribute's content and the associated consensus scores. Designers can select an attribute by dragging the corresponding puzzle piece from each dimension into the right-side puzzle area, where they can construct a comprehensive design entity that spans all design space dimensions (\autoref{fig:designer interface}-C1). Below the puzzle area, a templated textual prompt is displayed, generated from the design space, and serves as a descriptive representation of the design. When an attribute is selected, the relevant section of the template text is automatically populated with the corresponding attribute details.

\par
\begin{figure}[h]
    \centering
    \includegraphics[width=\linewidth]{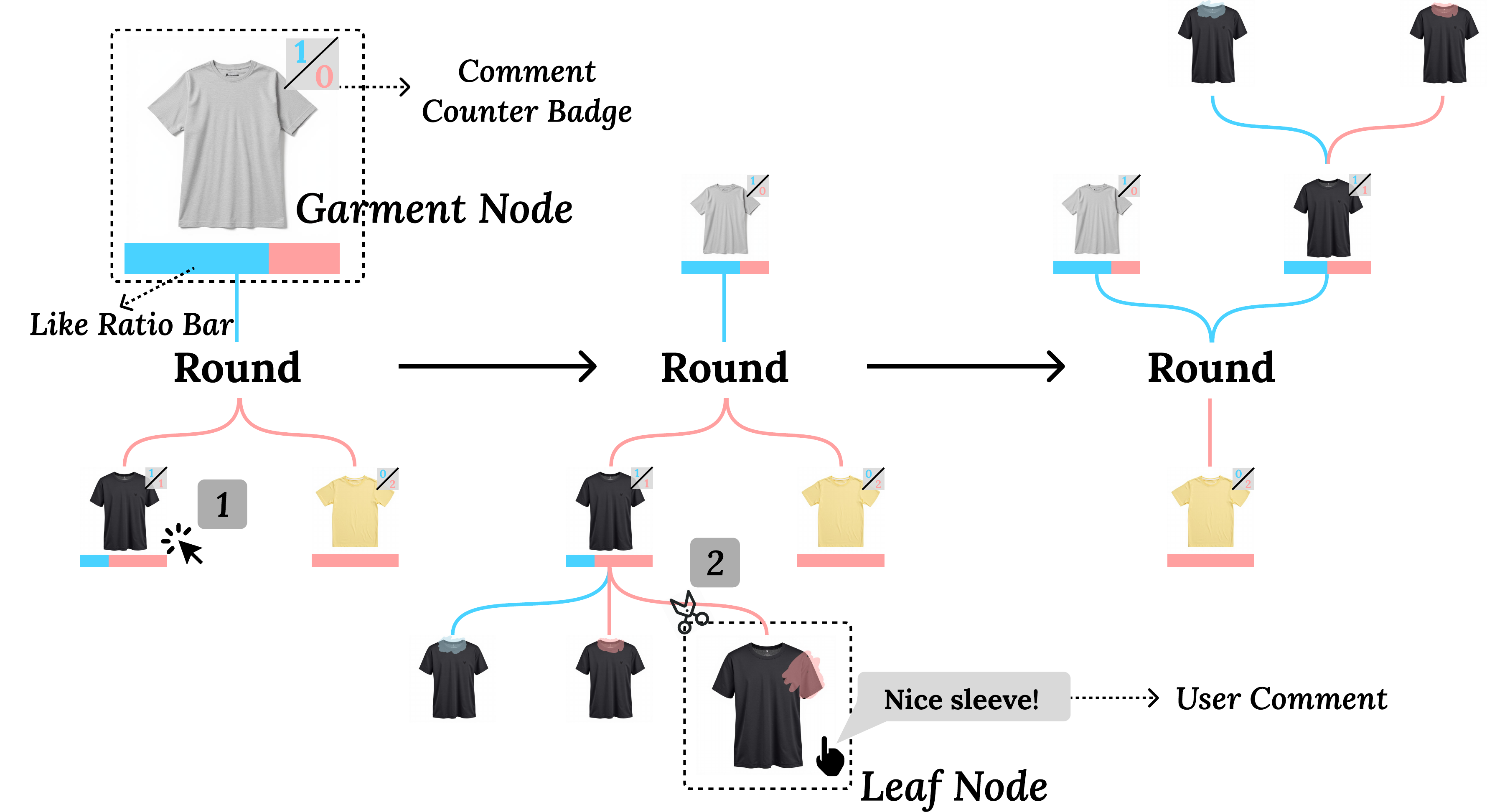}
    \caption{The detailed explanation of the tree structure and its interactions for an attribute. Designers can click a garment node in the first layer to view detailed user feedback in the second layer, showing interactions and comments. Designs with an equal or lower like-to-dislike ratio are classified as "dislike" and displayed above the root node.}
    \label{fig:pruning}
\end{figure}

 When a designer clicks on a puzzle piece representing an attribute within a design space dimension, a hierarchical tree structure for that attribute is displayed below (\autoref{fig:designer interface}-C2). The root node represents the attribute, with garments associated with the attribute appearing as child nodes in the first layer. As shown in detail in \autoref{fig:pruning}, these child nodes are sorted into two groups based on users' overall feedback—positive nodes are placed at the top, while negative nodes are shown at the bottom. The \textit{Comment Counter Badge} in the upper-right corner indicates the number of comments from overall like or dislike user feedback. Below each node, a \textit{Like Ratio Bar} visualizes the proportion of likes (red) and dislikes (blue) the current design received in user feedback. By clicking on any garment node, the designer can expand it to reveal the leaf nodes, which contain the interaction results for that garment within the current dimension. These results include any user-brushed areas and related comments. Designers can review this data and refine the tree by clicking on connecting lines to exclude specific user interactions or garments from the fine-tuning process for subsequent garment generation. For example, if a designer notices that a user-marked area related to the sleeve is mistakenly categorized under the collar due to incorrect attribute selection, they can exclude this from the fine-tune process. Similarly, if excluding certain user interactions causes a garment's positive feedback score to drop below the negative score, it will be excluded from further refinement.

\par Once designers have completed their review and interactions, they can click the \raisebox{-0.4ex}{\includegraphics[height=2.3ex]{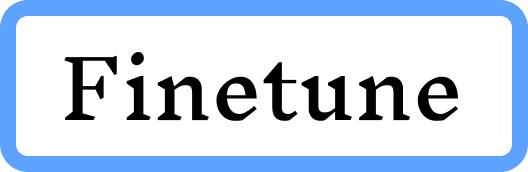}} at the top of the tree structure to refine the retained images. The system then uses these images, along with their corresponding attributes, as data to fine-tune the generative model, enabling designers to integrate their preferences into the generation process. A progress bar next to the button indicates the fine-tuning progress. After the fine-tuning is complete, designers can specify the number of design images to generate and click the \raisebox{-0.4ex}{\includegraphics[height=2.3ex]{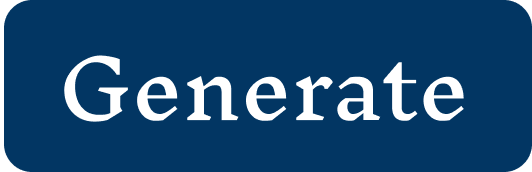}} to produce the final designs.

\subsubsection{Design generation with informed user feedback}
\par The \textit{Informed Generation View} (\autoref{fig:designer interface}-B) displays the generated design images. If multiple images are generated, users can navigate between them using the arrows on either side. To assist designers in analyzing the generated results alongside user feedback, the system also presents predicted user preference scores for each design, ranging from \textit{dislike} to \textit{like} (0-1). For each user's prediction, the system utilizes \textit{SHapley Additive exPlanations (SHAP)}~\cite{lundberg2017unified}, with dimensions from the design space serving as features to illustrate each dimension's contribution to the predicted preference score. Each dot within a dimension represents the contribution of that feature to an individual user's predicted preference. Additionally, the bar chart in each dimension displays the average contribution value of that feature across all users. To further enhance the designer's understanding of individual user feedback, clicking on a specific dot highlights the user's contribution values across all dimensions through connecting lines. A summary with an explanatory text is also provided below, offering an intuitive interpretation of the user's preference analysis. Once the designer is satisfied with a design, they can click the \raisebox{-0.4ex}{\includegraphics[height=2.3ex]{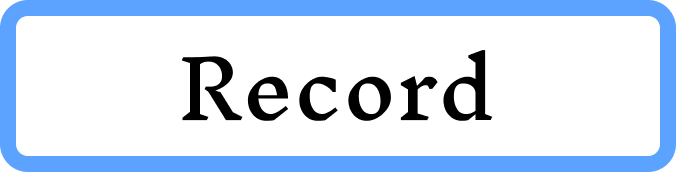}} to save the design information into the \textit{Design Library} for use in subsequent iterations.

\section{TECHNICAL DETAILS}

\par \textit{DesignBridge} employs \textit{Vue.js}\footnote{\url{https://vuejs.org}} for front-end development and utilizes \textit{Python} for the back-end implementation. The frontend and backend communicate via the \textit{Flask}\footnote{\url{https://flask.palletsprojects.com}} framework. For generative model workflows, we adopt \textit{ComfyUI}\footnote{\url{https://github.com/comfyanonymous/ComfyUI}} to support local development. Further technical implementation details are provided in the following section.

\subsection{Initial Design Framing}
\subsubsection{Initial Intent Processing}
\par To transform users' high-level requirements regarding \textit{Scene}, \textit{Type}, and \textit{Principle} into intuitive representations and a concrete design language, we employ the \textit{GPT-4o API} for semantic analysis and interpretation. The \textit{Scene} input is first expanded into a detailed scene description, which is then used in conjunction with \textit{ComfyUI} and the \textit{VisionRealistic v2 FluxDev} model to produce high-quality scene graphs tailored to try-on scenarios. Subsequently, the inputs from \textit{Scene}, \textit{Type}, and \textit{Principle}, along with a user-defined filtering level specified via the \textit{filter slider}, are used to exclude a subset of attributes from the full design space. Detailed prompts for scene description generation and attribute filtering can be found in \autoref{sec:Prompt-tryon}, \autoref{sec:Prompt-filter}. 

\subsubsection{Database Construction for Preference Expression}
\label{sec:512}
\par The task begins with the construction of a comprehensive garment dataset sourced from \textit{Farfetch}, a fashion e-commerce platform. Each data instance \( e \) comprises an image (\( I \)), a set of textual descriptions (\( T = \{t_1, t_2, \dots, t_m\} \)), and a set of attributes spanning nine dimensions of the design space (\( A = \{a_1, a_2, \dots, a_9\} \)). The image (\( I \)) and text descriptions (\( T \)) are obtained as described in \autoref{sec:34}. Attribute values (\( A \)) are first extracted from the textual descriptions \( T \) using a rule-based approach and subsequently completed for missing dimensions via the GPT-4o API. To ensure annotation quality and semantic correctness, the designers (P1-P3) from the formative study are actively involved throughout the labeling process. The prompt used for image feature identification is provided in \autoref{sec:Prompt-stage1}.

\par This curated garment dataset is then used to fine-tune a \textit{LoRA} model~\cite{hu2022lora} built on top of the pre-trained \textit{Vision Realistic Flux} model. The goal is to teach the generative model to internalize and reproduce visual characteristics aligned with the attribute dimensions \( A \). For each data instance \( e \), the corresponding image (\( I \)) is resized to \( 768 \times 768 \) pixels and normalized. A prompt \( P \), created by concatenating the attribute values \( A = \{a_1, a_2, \dots, a_9\} \) into a descriptive textual format and appending the trigger word ``real garment'', is used to guide domain-specific generation. The LoRA model employs a low-rank adaptation with rank \( r = 64 \), a learning rate of \( 4 \times 10^{-4} \), and is trained over 1500 steps with a batch size of 1. Training is performed on an NVIDIA RTX 4080 SUPER GPU, requiring approximately 11 hours. After fine-tuning, we construct the design database by randomly sampling attribute combinations from the design space, forming prompts, and generating corresponding images using the refined model.

\subsection{Preference Expression Collection}
\subsubsection{Virtual Try-On}

\par Virtual try-on technology supports both designer database reviews and user preference elicitation. We detail the latter as a representative workflow, which involves \textit{mannequin generation} and \textit{clothing replacement}. First, the system synthesizes user attributes (\textit{gender}, \textit{height}, \textit{weight}) into a textual prompt for the \textit{VisionRealistic v2 FluxDev} model to generate a mannequin portrait (\autoref{sec:Prompt-avatar}). Following this, \textit{LayerMask}\footnote{\url{https://github.com/chflame163/ComfyUI_LayerStyle_Advance}} isolates the human region via segmentation, and the background is replaced with a designer-provided scene for contextual alignment. During clothing replacement, a \textit{LayerStyle} segmenter extracts the primary garment layer from the recommended clothing image. This layer is then integrated onto the mannequin using the \textit{CatVTON Wrapper}~\cite{chong2024catvton} virtual try-on module, ensuring a natural fit and visual consistency with the mannequin's body shape and the customized scene.

\subsubsection{Local Preference Hypothesis}
\par To enable rapid and accurate preference elicitation, user-brushed regions must be efficiently translated into intent signals across specific design dimensions. The system masks out non-brushed areas, retaining only user-selected regions for analysis. These regions are processed with \textit{FashionCLIP}~\cite{chia2022contrastive}, a vision–language model fine-tuned for the fashion domain that extracts feature vectors representing visual attributes such as texture, shape, and color. The extracted features are then fed into a convolutional neural network (CNN) that produces nine parallel outputs, each corresponding to one design space dimension. The model is trained on 1,380 multi-labeled brush images to learn accurate mappings between visual cues and design intent.

\subsubsection{Personalized Preference Modeling with Recommendation Algorithms}
\par To model personalized preferences, our recommendation framework integrates attribute-based and visual features. Each item is represented by a hybrid feature vector: a 51-dimensional one-hot encoding (mapping to the nine-dimensional design space) concatenated with a 50-dimensional visual vector. The latter is derived from a pre-trained \textit{ResNet-50} 1000-dimensional embedding, reduced via an end-to-end trained fully connected layer. A Personalized Preference Neural Network (PPNN) is trained incrementally at each iteration using these hybrid vectors and binary user feedback (1 for liked, 0 for disliked). The model predicts preference probabilities for unseen items to enable adaptive recommendations.

\par During the cold-start phase, the system initializes user modeling using a designer-curated seed ranking by recommending 10 diverse items to bootstrap the preference model. In each subsequent iteration, 5 new items are recommended. At iteration \( t \), the system selects items that maximize information gain based on entropy, which is calculated for each item \( j \) as:
\[
h(\mathbf{x}_j) = - \left( p_j^{(t)} \log_2 p_j^{(t)} + (1 - p_j^{(t)}) \log_2 (1 - p_j^{(t)}) \right),
\]
where \( p_j^{(t)} \) denotes the predicted probability of user preference. The top-5 items with the highest entropy scores are selected to promote exploration across the design space, thereby refining the model through continuous user feedback.

\subsection{Preference-Integrated Design}
\subsubsection{Attributes Consensus Score}
\par Once all users' preferences are fully collected, the system computes the User Preference Utility for each user \( i \) on a given attribute \( a \), defined as:
\( \text{UPU}_{i,a} = \frac{L_{i,a} + 1}{L_{i,a} + D_{i,a} + 2} \).
Here, \( L_{i,a} \) and \( D_{i,a} \) represent the number of times user \( i \) has liked or disliked attribute \( a \), respectively, based on their brush interactions. Laplace smoothing is applied to account for limited data and prevent zero-probability issues. To capture the overall group preference for each attribute, the Attribute Consensus Score for each attribute \( a \) is computed as the geometric mean of the individual user utilities:
\( \text{ACS}_a = \left( \prod\limits_{i=1}^{n} \text{UPU}_{i,a} \right)^{\frac{1}{n}} \). This score is then normalized to the range \( (0, 1) \) to facilitate consistent comparison across attributes.

\subsubsection{Integrating User Preferences into the Generative Model}
\label{sec:532}
\par Inspired by the \textit{``Keep''} mechanism from \textit{IntentTuner}~\cite{zeng2024intenttuner}, we introduce a \textit{Preference-Guided Fine-Tuning} approach that leverages brush-marked data from the \textit{Preference Expression Collection} stage to further align the generative model with user preferences while preserving garment semantics. The brush-marked data is organized at the attribute level, allowing designers to selectively exclude specific garments or user annotations. For fine-tuning, only data with a user ``like'' ratio exceeding 0.5 is retained.

\par For each brushed region \( B_i \) on a garment image \( I \), the system generates a binary mask \( M_i \) based on the brush stroke coordinates \( (x_{\text{min}}, y_{\text{min}}, x_{\text{max}}, y_{\text{max}}) \), where \( M_i(x, y) = 1 \) within the region \( B_i \) and 0 elsewhere. The fine-tuning builds on the \textit{VisionRealistic v2 FluxDev} model previously trained during the \textit{Initial Design Framing} phase. A dedicated \textit{LoRA} module is trained for each attribute in the nine-dimensional design space, employing a Local Perceptual Loss strategy to enhance attribute-level fidelity:
\[
\mathcal{L}_{\text{local}, a_i} = \sum_{j} || F_{\theta}(I_j \cdot M_j) - F_{\theta}(T_{a_i}) ||^2.
\]
Here, \( F_{\theta} \) denotes the feature extractor from the U-Net backbone; \( I_j \cdot M_j \) is the masked image region corresponding to attribute \( a_i \); and \( T_{a_i} \) represents the CLIP text embedding of the attribute. To balance global coherence and localized control, the training objective combines this loss with a CLIP-based guidance loss: 
\[
\mathcal{L}_{a_i} = \lambda_{\text{CLIP}} \mathcal{L}_{\text{CLIP}} + \lambda_{\text{local}} \mathcal{L}_{\text{local}, a_i}
\]
with weighting parameters empirically set to \( \lambda_{\text{CLIP}} = 0.6 \) and \( \lambda_{\text{local}} = 0.4 \). All training is performed on an NVIDIA RTX 4080 SUPER GPU, with each attribute-specific fine-tuning process requiring approximately 12 minutes.

\subsubsection{User Preference Analysis}
\par To gain deeper insights into user preferences, we analyze the PPNN trained for each user. We begin by extracting the outputs from the Logits Layer of the PPNN, which represent the model's confidence scores before the final binary classification. These logits reflect the predicted strength of a user's preference for each clothing item. Next, we assess the influence of nine specific attributes pre-selected by designers from the full nine-dimensional design space on the model's predictions. To do this, we apply the Shapley Value~\cite{roth1988shapley}, a game-theoretic approach that quantifies the marginal contribution of each attribute to the prediction outcome. This allows us to identify which attributes most significantly influence the model's decisions. The resulting analysis provides interpretable, attribute-level insights into the factors driving individual user preferences, highlighting the relative importance of designer-selected attributes in shaping recommendation outcomes.

\subsection{Technical Evaluation}
\label{technical evaluation}

\par Our technical evaluation focuses on the two-stage fine-tuning pipeline, the core technical contribution of \textit{DesignBridge}. Other components, such as interpretability features and preference collection via recommendation, were included to complete the workflow and enhance user experience. These techniques are well established in prior work, and their effectiveness in our context is further examined in the user study. The two fine-tuning stages serve complementary purposes: the first, conducted in the Initial Design Framing phase (\autoref{sec:512}), enriches the pre-trained model’s understanding of design space semantics; the second, implemented in the Preference-Integrated Design stage (\autoref{sec:532}), incorporates user preferences into the generative process to achieve personalized design outcomes. To validate the effectiveness of this pipeline, we conducted a technical evaluation assessing the contribution of each fine-tuning stage to the overall generative performance.

\subsubsection{Baseline}
\par We selected the pre-trained \textit{VisionRealistic v2 FluxDev} model as a baseline, as its photorealistic rendering capabilities aligns closely with the fashion design domain's demand for high-fidelity garment visualization. This model also served as the foundation for our fine-tuning process. Additionally, we included GPT-4o, which has demonstrated state-of-the-art performance in image synthesis. However, because GPT-4o does not support LoRA fine-tuning, our evaluation was limited to assessing its generative outputs, and no significance testing was conducted for this baseline.


\subsubsection{Procedure}
\begin{figure}[h]
    \centering
    \includegraphics[width=\linewidth]{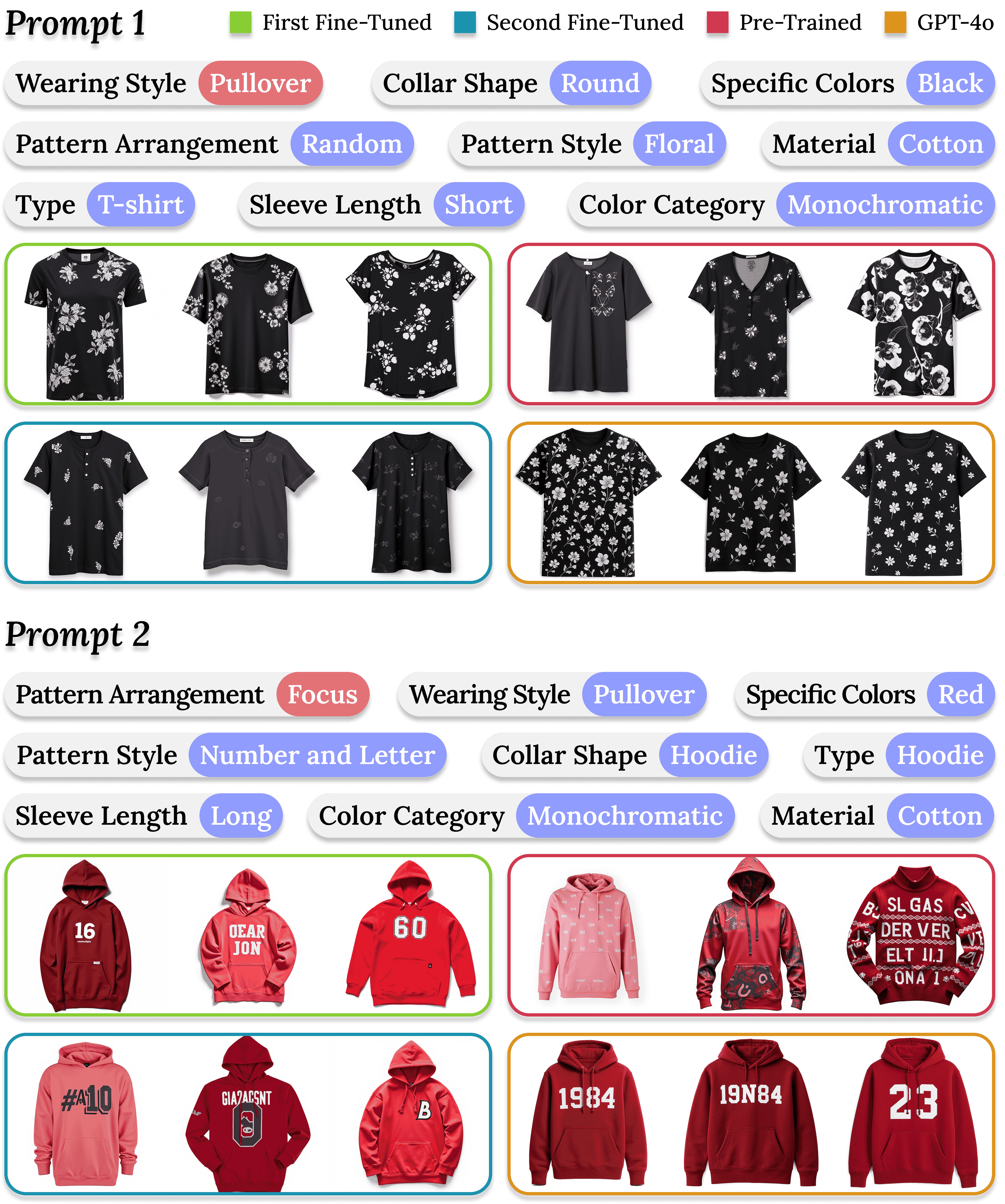}
    \caption{Two example combinations of prompt and target attribute, each shown with images generated by four different models. By comparison, our fine-tuned models comprehensively outperform the pre-trained model in Design Space Capture. Notably, the second fine-tuned model better captures user preferences for button styles. GPT-4o produces impressively high-quality images, but lacks the ability to generate design diversity.}
    \label{fig:technicalprompt}
\end{figure}

\par Following IRB approval, we recruited 3 professional designers, each with over three years of fashion design experience, along with 9 target users selected by the designers. To ensure a shared understanding, we first provided all participants with a detailed explanation of the design space and the system's operational principles. We then crafted 9 design prompts by sampling attribute combinations from the nine-dimensional design space. Each prompt was reviewed and approved by the designers to ensure semantic coherence. For every prompt, the designers selected an additional attribute, based on which we retrieved 50 images from the expert-annotated Farfetch dataset. To avoid potential bias that might occur when using outputs from the first fine-tuning to evaluate the second, we used Farfetch images instead. This ensured that the effectiveness of each model stage could be independently assessed in the technical evaluation. Users then provided ``like'' or ``dislike'' brush feedback focused solely on the selected attribute across these images, and this feedback was used to inform the second fine-tuning stage.

\par Each design prompt was used to generate three images from each of the four models under evaluation. To evaluate these images, we defined two assessment dimensions: \textit{Design Space Capture}, which includes attribute accuracy (M1) and design consistency (M2), and \textit{User Preference Integration}, which covers satisfaction with the target attribute (M3) and the overall design (M4). The first dimension was evaluated by designers, while the second was assessed by users. All ratings were collected on a 7-point Likert scale and reflect participants’ subjective judgments. We used the Wilcoxon signed-rank test to examine the results across the four models and assess the significance of their differences. To promote focus and consistency, participants were instructed to evaluate each image within 15 seconds, concentrating specifically on design-relevant visual features rather than general image quality. All participants contributed to the scoring process. Additionally, we conducted follow-up interviews with eight participants (two designers and six users) to collect deeper insights into their perceptions of the generated designs.

\subsubsection{Result}
\par
\begin{figure*}[h]
    \centering
    \includegraphics[width=\textwidth]{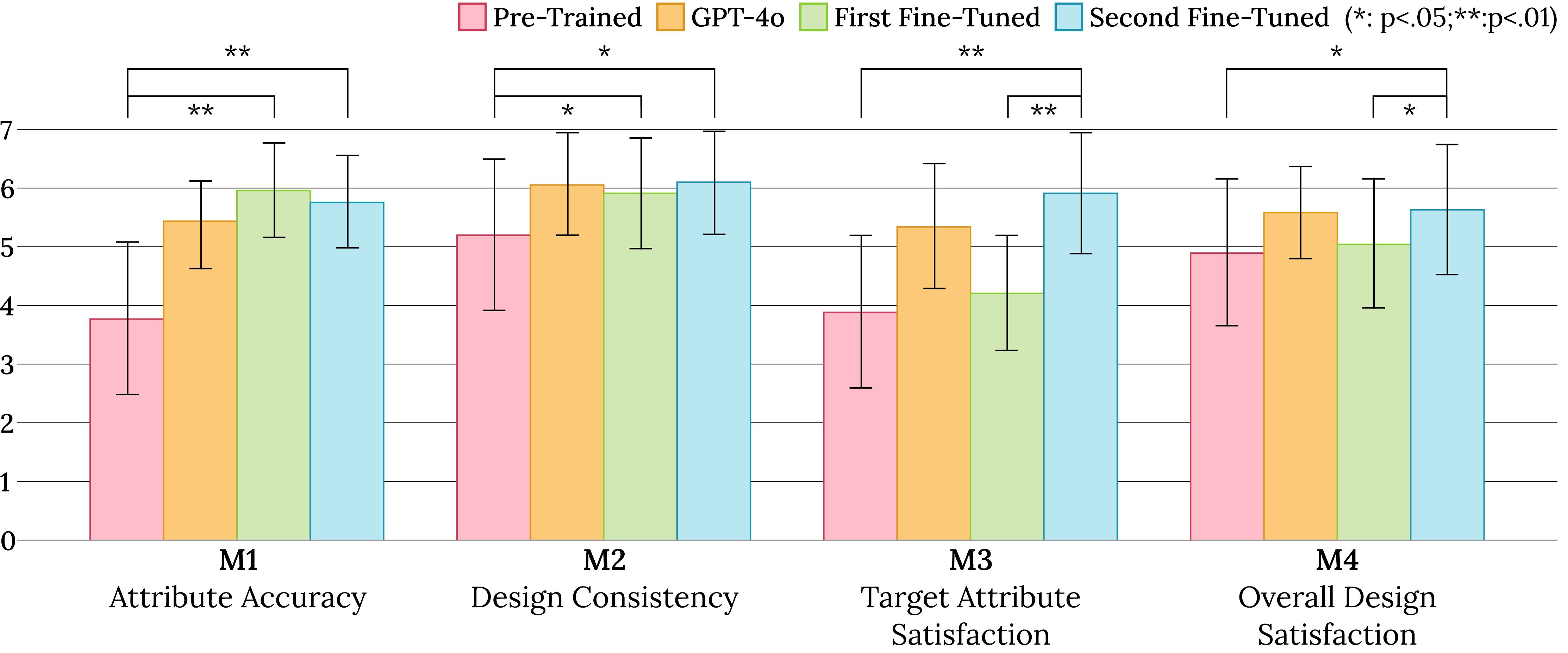}
    \caption{Results of the subjective experiment comparing Design Space Capture and User Preference Integration across our fine-tuned models and two baseline models.}
    \label{fig:technicalresult}
\end{figure*}

 We calculated the average scores for each evaluation metric across all images generated by each model. No significant outliers were observed in the ratings. As illustrated in~\autoref{fig:technicalresult}, both of our fine-tuned models consistently outperformed the pre-trained \textit{VisionRealistic v2 FluxDev} model across all metrics, with especially notable improvements in M1 (attribute accuracy) and M2 (design consistency). These results suggest that our fine-tuning strategies significantly enhanced the models' ability to interpret and convey the intended design semantics. For instance, under Prompt 2 (\autoref{fig:technicalprompt}), which emphasized a ``focus'' pattern layout, the pre-trained model generated outputs with arbitrary pattern positioning, whereas both fine-tuned models accurately captured and expressed the specified arrangement.

\par The second fine-tuned model achieved a substantial improvement on M3 (satisfaction with the target attribute) and M4 (overall design satisfaction), outperforming all other models. To better understand these results, we conducted follow-up interviews. Most participants (5/6) expressed a clear preference for a “buttoned round-neck black t-shirt” in pullover styles, disfavoring plain or zippered alternatives. One user (1/6) expressed a neutral view, noting that “these wear styles are about equally appealing.” For Prompt 1 (\autoref{fig:technicalprompt}), the second fine-tuned model successfully captured this preference trend, generating designs that aligned closely with majority tastes.  

\par GPT-4o demonstrated strong performance on M1 and M2, indicating its ability to align well with prompt-based instructions. However, several participants highlighted its lack of diversity. As one designer noted, ``\textit{We tried the same prompt on its official site multiple times; every image looked nearly identical, with no variation in color brightness, details, or even digit styles—sometimes two runs in the same window produced nearly the same result.}'' This perceived uniformity poses a limitation for the \textit{DesignBridge} workflow, which depends on visual diversity to support creative iteration in fashion design. Nevertheless, GPT-4o maintained competitive performance on M3 and M4, likely attributable to its high overall aesthetic quality and visual coherence. A notable trade-off, however, was its higher generation latency compared to both our fine-tuned models and the original pre-trained model.

\section{USER STUDY}
\par The technical evaluation presented in \autoref{technical evaluation} highlights the effectiveness of \textit{DesignBridge} in interpreting user intentions and incorporating user preferences throughout the design generation process. To further assess the system’s overall effectiveness and usability, we also conducted a between-subjects study ($N$=42).

\subsection{Methodology}
\subsubsection{Participants}

\par Following IRB approval, our study began by recruiting ten designers (D1–D10) through our collaboration team. Each designer selected an individual design theme from a set of predefined options based on their professional expertise and creative interests. After the themes were confirmed, we recruited thirty-two users (U1–U32) via a mailing list and word of mouth, pairing each designer with several users whose preferences aligned with the corresponding design theme. All designers had over three years of professional experience in fashion design and prior involvement in co-design practices with users. Similarly, all user participants reported previous experience engaging in co-design activities with designers within the fashion domain. Detailed participant information is provided in \autoref{sec:participants-userstudy}.

\subsubsection{Baseline} 
\par For comparison, we developed a baseline system that preserved the overall workflow of \textit{DesignBridge}, including the three main stages—\textbf{Initial Design Framing}, \textbf{Preference Expression Collection}, and \textbf{Preference-Integrated Design}—while deliberately omitting several key functions.

\par In the \textbf{Initial Design Framing} stage, designers described their design requirements in natural language. The system parsed these inputs and used the pre-trained \textit{VisionRealistic v2 FluxDev} model to generate an initial set of design images. Unlike \textit{DesignBridge}, this version did not construct a structured design framework or filter incompatible attributes from the design space.

\par During the \textbf{Preference Expression Collection} stage, users were presented with garment images determined by the designer, either following a predefined sequence or through manual selection, to express their preferences. The virtual try-on function was retained to support contextual visualization during interaction. Users could still provide both brush-based and textual feedback on specific garment regions; however, the system did not classify these inputs as positive or negative, nor did it perform semantic interpretation. Additionally, the recommendation-based image allocation mechanism available in \textit{DesignBridge} was removed.

\par In the \textbf{Preference-Integrated Design} stage, designers were provided with a simplified table view summarizing user interactions. Each row began with a garment image that had received feedback, followed by thumbnails representing individual users' brush and text comments. Clicking on a thumbnail allowed designers to inspect detailed interaction records. A prompt input panel below the table was pre-filled with a default generation instruction, which designers could modify to specify custom prompts and the number of images to produce in the next iteration. Compared with \textit{DesignBridge}, this baseline omitted several mechanisms, including recommendation-based preference elicitation, fine-grained semantic scoring within the design space, and user preference modeling for semantics-specific fine-tuning.

\par Subsequently, designers reviewed the generated design results and selected garments they found satisfactory. These selections could be carried forward into future iterations, accompanied by optional textual notes for recording design rationales or observations. Unlike \textit{DesignBridge}, the baseline did not incorporate predictive analysis or interpretability features for user preferences. Designers instead relied solely on their own judgment to determine which designs to refine or regenerate

\par Overall, this baseline was constructed to ensure a fair comparison with \textit{DesignBridge} while isolating the effects of its core components. By maintaining an equivalent workflow but reducing functionality, it allows us to assess the specific contributions of \textit{DesignBridge}'s key mechanisms to the collaborative design process.

\begin{figure*}[t]
    \centering
    \includegraphics[width=\textwidth]{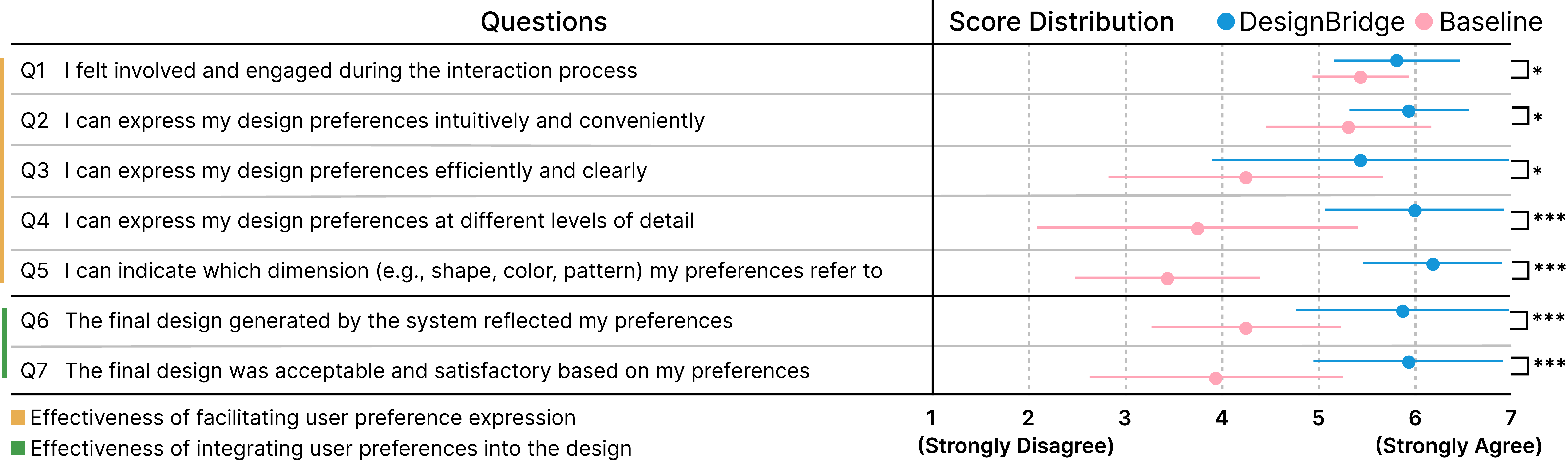}
    \caption{The effectiveness of \textit{DesignBridge} for users.}
    \label{fig:effectiveness user}
\end{figure*}

\subsubsection{Procedure}
\par We developed a dedicated co-design task centered on garment design within the fashion domain, aligned with established industry practices. In this task, each designer first selected a design theme from a set of predefined options, and then created an initial design framework, collaborated with users to gather preferences and requirements, analyzed the feedback, and iteratively refined the design. Participants were organized into ten designer–user groups. We then assigned five groups to the \textit{DesignBridge} condition and five to the baseline condition. Each group consisted of one designer and three to four users, completing the co-design task independently.
\par Before the study, we provided participants with a brief overview of the research context and collected their demographic information. All procedures were conducted with informed consent. Audio and screen recordings were captured throughout the study for subsequent analysis.
\par We began by introducing the overall system workflow shared by both the \textit{DesignBridge} and baseline conditions. Participants were then separated into two roles—designers and users—and each received a role-specific walkthrough. Designers were introduced to the design interface corresponding to their assigned condition, while users were guided through the user interface. Concrete examples were used to demonstrate the system's functionality, and participants were given time to freely explore the system. We addressed their questions to ensure a clear understanding of the system's features and workflow. This orientation phase lasted approximately $20$ minutes.
\par During the task, designers first constructed an initial design framework, after which users interacted with the system to express their preferences. Once user feedback had been collected, designers analyzed the results and generated updated design outputs. Both conditions followed the same iterative workflow, allowing designers to refine and regenerate designs based on user input until a satisfactory outcome was achieved. The entire process took approximately two and a half hours. Each designer received $25$ per hour and each user received $10$ per hour as compensation for their time and contribution.

\subsubsection{Measurement}
\par Upon completing their interactions with the user interface, users provided feedback on their experience with the system via an in-task questionnaire. After designers finalized their design exploration in the design interface, users were able to review the generated designs and complete a post-task questionnaire to share their opinions on the designs. Meanwhile, designers completed a post-task questionnaire to provide feedback on the co-design process. The questionnaire design primarily focused on capturing the perspectives of designers and users regarding the system's effectiveness and usability. For users, the questionnaire addressed the system's effectiveness in facilitating user preference expression, its effectiveness in integrating user preferences into the design, and its usability. For designers, the questionnaire evaluated the system's effectiveness in supporting the collection and analysis of user feedback, its effectiveness in collecting and analyzing user preferences, the effectiveness of design adjustment and iteration, and its usability. We employed the \textit{Mann–Whitney U test} to analyze the data between the baseline and \textit{DesignBridge} conditions and to examine the significance of their differences. We also conducted semi-structured interviews with participants to obtain qualitative feedback regarding the system.

\subsection{Results and Findings}

\begin{figure*}[h]
    \centering
    \includegraphics[width=\linewidth]{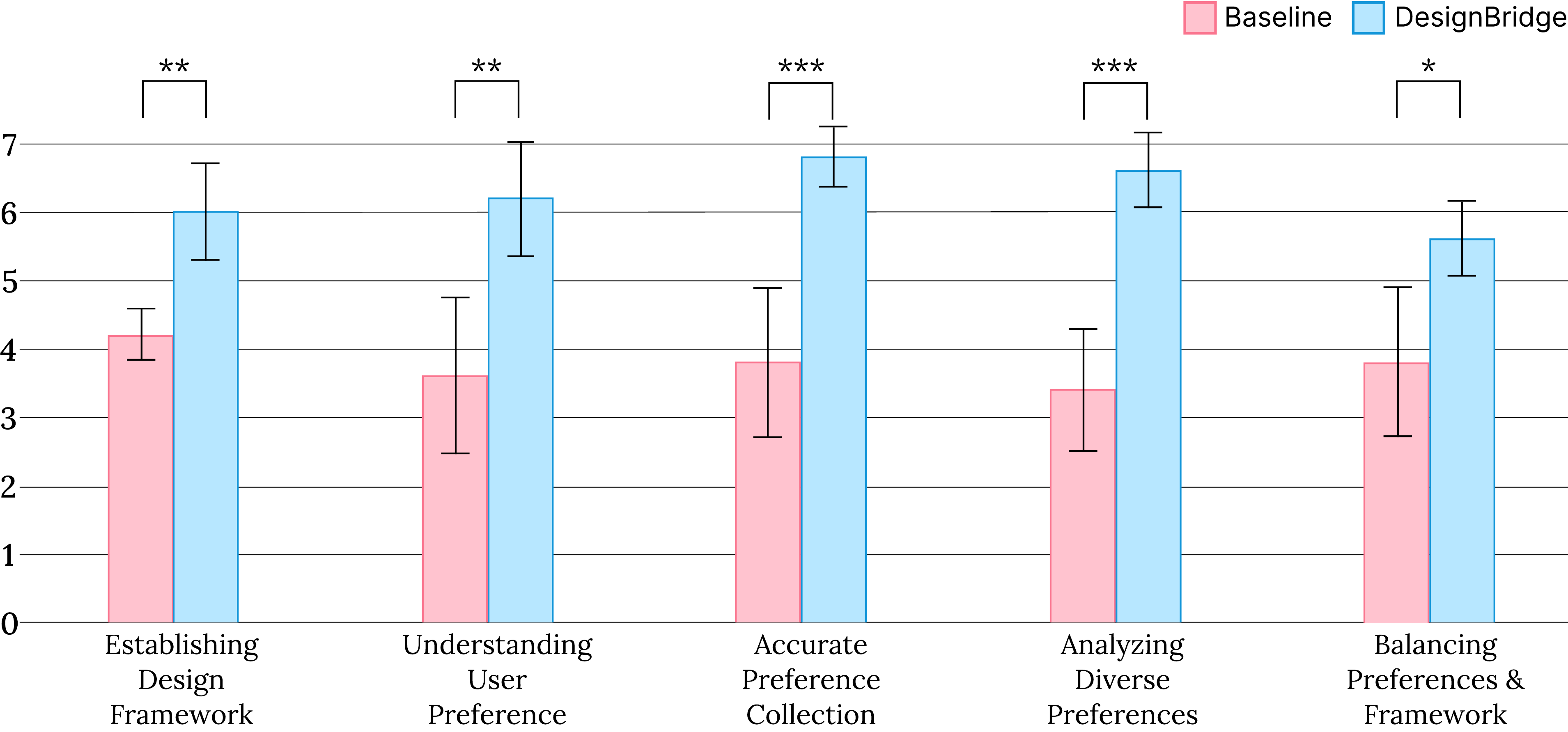}
    \caption{The effectiveness of collecting and analyzing user preferences on Baseline and \textit{DesignBridge}.}
    \label{fig:effectiveness designer 1}
\end{figure*}

\begin{figure}[h]
    \centering
    \includegraphics[width=\linewidth]{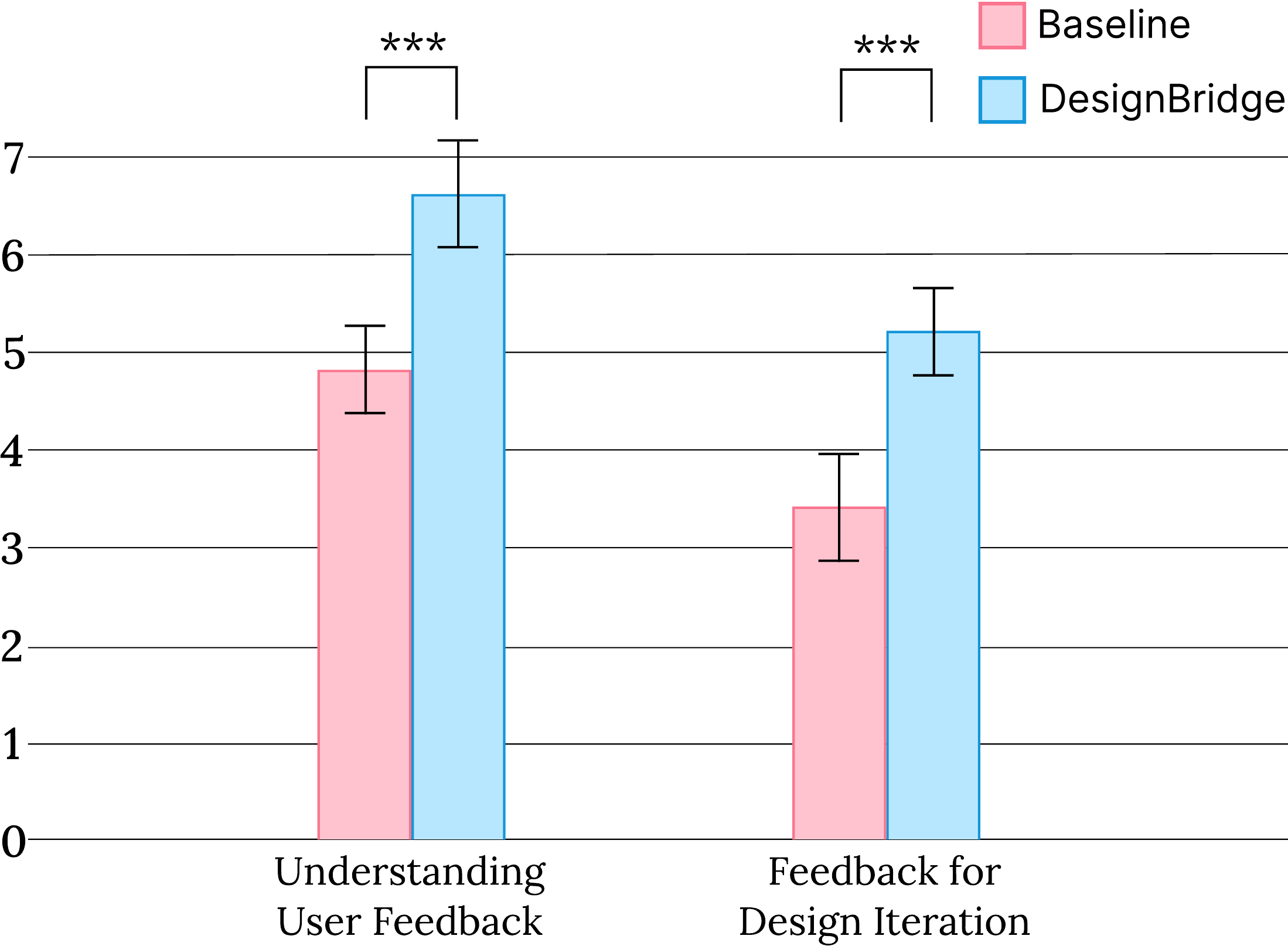}
    \caption{The effectiveness of design adjustment and iteration on Baseline and \textit{DesignBridge}.}
    \label{fig:effectiveness designer 2}
\end{figure}

\subsubsection{Effectiveness of facilitating user preference expression [User]}
\par \autoref{fig:effectiveness user} presents user feedback on the effectiveness of the baseline and \textit{DesignBridge} system in facilitating preference expression. Users generally acknowledged \textit{DesignBridge}'s support for engagement within design contexts, reporting that it fostered more active and sustained participation compared to the baseline system (Q1). As U3 noted, ``\textit{Through scene illustrations and virtual try-ons, the system helps me intuitively understand what we are doing, which makes it easier to engage in the whole process.}'' In contrast, several users described the baseline as ``\textit{less immersive}'' and ``\textit{hard to stay focused on}''. All participants agreed that the system provided a straightforward and intuitive way to express preferences, while the baseline often required more explanation or abstract description (Q2). As U10 remarked, ``\textit{Compared to previous approaches, this feels much easier and more natural—I don't need to say a lot. I can just express what I have in mind based on my intuitive thoughts.}'' While U19 commented that in the baseline condition, she ``\textit{struggled to find the right words}'' and felt ``\textit{uncertain if the designer understood what [she] meant.}'' Regarding the preference articulation, most of the users also reported that \textit{DesignBridge} system enabled clear and efficient expression of their preferences compared with baseline (Q3). U4 expressed the need for repeated brushing when selecting local regions. We are considering adding interaction methods such as lasso selection or bounding boxes for more effectively capturing areas. Moreover, participants recognized the system's ability to support fine-grained preference expression, including both global and local (Q4), and the design dimensions they refer to (Q5). However, the baseline was reported only allowed participants to express general preferences, making it difficult to capture nuanced details. U9 acknowledged \textit{DesignBridge}'s ability to capture both global and local preferences, emphasizing that ``\textit{clothing preferences can be complex and layered—just because I didn't like the overall design doesn't mean I didn't like certain parts of it, and the system picks up on that.}'' U2 appreciated the dimension-specific expression, stating that ``\textit{Just because appreciating a certain part doesn't mean I like everything about it—that's not how people usually think.}'' and acknowledged the system for allowing ``\textit{quick and easy ways to share what I think about specific aspects.}''

\begin{table*}[h]
  \centering
  \caption{Statistical usability feedback for \textit{DesignBridge} from designers and users
  (significance: * $p<.050$, ** $p<.010$, *** $p<.001$).}
  \label{tab:usability}
  \setlength{\tabcolsep}{6pt}
  \renewcommand{\arraystretch}{1.18}
  \begin{tabular}{l l c c S[scientific-notation = true, round-mode=places, round-precision=2, round-pad, table-format=1.2e-1] c}
    \toprule
    \textbf{Group} & \textbf{Dimension} &
    \textbf{DesignBridge} & \textbf{Baseline} &
    {\textbf{$p$-value}} & \textbf{Sig.} \\
    & & \textbf{Mean/S.D.} & \textbf{Mean/S.D.} & & \\
    \midrule
    \multirow{5}{*}{Designer}
      & Frequency of use     & 6.60/0.55 & 3.60/0.89 & 0.000220 & \textbf{***} \\
      & Ease of use    & 6.20/0.42 & 5.48/0.30 & 0.014957 & \textbf{*}  \\
      & Functionality  & 6.10/0.22 & 3.90/0.74 & 0.000210 & \textbf{***} \\
      & Confidence     & 6.40/0.55 & 4.40/0.55 & 0.000417 & \textbf{***} \\
      & Learnability   & 6.00/0.35 & 5.60/0.42 & 0.141113 & - \\
    \midrule
    \multirow{5}{*}{User}
      & Frequency of use     & 5.88/1.09 & 4.63/1.09 & 0.002845  & \textbf{**} \\
      & Ease of use    & 5.90/0.72 & 5.65/0.44 & 0.243542  & -  \\
      & Functionality  & 5.75/1.00 & 4.63/0.81 & 0.001464  & \textbf{**} \\
      & Confidence     & 6.38/0.50 & 5.13/0.81 & 0.0000108 & \textbf{***} \\
      & Learnability   & 6.28/0.26 & 6.13/0.47 & 0.248711  & - \\
    \bottomrule
  \end{tabular}
\end{table*}

\subsubsection{Effectiveness of integrating user preferences into the design [User]}
\par In \autoref{fig:effectiveness user}, we also present the user feedback on the effectiveness of integrating use preferences into the design. Users generally acknowledged that the designer-side generated designs in \textit{DesignBridge} better reflected their expressed preferences compared to the baseline system (Q6). As U5 noted, ``\textit{In earlier interactions, I expressed a lack of appealing patterns in the presented designs. I now see generated images with elements like numbers and letters, which better align with my preferences.}'' Additionally, nearly all participants found the final designs to be generally acceptable and satisfactory (Q7), whereas the baseline designs were often described as ``less representative'' or ``too generic''. As U3 mentioned, ``\textit{This may not be exactly what I like most, but it's a result I can accept and recognize.}'' This is consistent with our design intent: rather than produce highly personalized results that perfectly satisfy a few users, the goal is to generate designs that accommodate the preferences of most users and are broadly acceptable. Participants emphasized that this outcome was much more apparent in \textit{DesignBridge} than in the baseline, where designs tended to converge toward designer biases rather than user-informed preferences.

\subsubsection{Effectiveness of collecting and analyzing user preferences [Designer]}
\par We also collected designers' feedback on the system's effectiveness (\autoref{fig:effectiveness designer 1}). Designers first acknowledged \textit{DesignBridge}'s support in helping them establish a design framework, while the baseline group has relatively low scores. Designer in \textit{DesignBridge} condition stating that ``\textit{this integrated approach makes the entire process, from inputting design information to generating initial design images, much more efficient}'' (D3). Regarding the collected user preferences, designers appreciated \textit{DesignBridge}'s ability to help translate user preferences into concrete design considerations. As D1 noted, ``\textit{I can directly see users' preferences for different design features such as color and pattern style, which saves me from having to sift through raw data to figure out what they want.}'' This contrasted with the baseline, where preference summaries were ``\textit{too abstract to guide specific design choices.}'' When analyzing feedback, designers recognized that \textit{DesignBridge}'s fine-grained data collection helped them better understand diverse user preferences and avoid ambiguity. D2 commented, ``\textit{In the tree diagram, I saw how users expressed their preferences regarding patterns. I found that the comments on disliked items either indicated a lack of patterns or strange ones. This inspired me to consider adding patterns like letters or numbers for decoration during generation.}'' D5 further emphasized that this ``\textit{helped designers better understand user preferences and their underlying reasons, supporting more informed design decisions.}'' While the feedback from the baseline group indicated that the preference is ``\textit{not concrete and accurate through users' raw descriptions}''.

\subsubsection{Effectiveness of design adjustment and iteration [Designer]} As illustrated in \autoref{fig:effectiveness designer 2}, designers valued the system's ability to support the comprehensive analysis of differing user preferences. D3 said, ``\textit{I noticed that a V-neck design had a roughly even split between users who liked it and those who didn't. After reviewing the feedback, I found that supporters appreciated the style, while someone felt it clashed with the short-sleeve format. So I pruned this design to avoid combining formal V-necks with casual short-sleeved garments.}'' Designers using the baseline reported that ``\textit{it was hard to identify the cause of disagreement without knowing their detailed preferences of different aspects.}'' (D9) Finally, designers acknowledged \textit{DesignBridge}'s role in helping them balance user preferences with the overall design framework. As one designer shared, ``\textit{I saw that both orange and purple received high user preference scores. Since the current design follows a summer casual theme, a warm color like orange might be less suitable. So, I ultimately chose purple.}'' This demonstrates that \textit{DesignBridge} supports nuanced trade-offs between user preferences and design intent, enabling designers to make more contextually appropriate design choices.

\par Regarding the generated design results, designers highlighted that our system, \textit{DesignBridge}, provided more actionable and interpretable insights than the baseline system, which only displayed predicted preference scores without detailed explanations. Designers acknowledged that \textit{DesignBridge}'s predicted user feedback helped them understand their preferences for the design. As D4 noted, ``\textit{When I get a new design, I'm eager to know how users might react to it. The system's predictions offer me an initial reference and guidance for making adjustments.}'' Additionally, designers found that detailed feedback is helpful for their design iterations, such as contribution scores of different design dimensions based on the prediction. D5 remarked, ``\textit{By examining the contribution scores of various design dimensions to user preferences, I can quickly identify which aspects lead users to like or dislike a particular design.}'' This level of interpretability was absent in the baseline, where designers had to ``\textit{manually infer which design features influenced user preferences, often through trial and error.}''

\subsubsection{Usability}

\par As shown in \autoref{tab:usability}. We evaluated the system's usability from five aspects: \textit{Frequency of use}, \textit{Ease of use}, \textit{Functionality}, \textit{Confidence}, and \textit{Learnability}. The results show that both designers and users rated \textit{DesignBridge} significantly higher than the baseline in most dimensions. Designers reported notably higher scores in \textit{Frequency of use}, \textit{Ease of use}, \textit{Functionality}, and \textit{Confidence} ($p<.05$), while users showed significant improvements in \textit{Frequency}, \textit{Functionality}, and \textit{Confidence} ($p<.01$). No significant differences were observed in \textit{Learnability} for either group.


\subsubsection{Suggestions}
\par During the study, we also collected several notable suggestions from participants. One frequently mentioned suggestion was to support more flexible and fine-grained expressions of preference combinations. As one user (U6) remarked: ``\textit{When I look at a clothing item, sometimes I feel that individual parts of the design are fine, but the way they are combined into a whole doesn't work.}'' This highlights the need to express preferences over combinations of attributes across different dimensions to better capture nuanced user intentions. Additionally, we noticed that users are experiencing a potential cold-start issue during the preference elicitation process. Initially, the system presents relatively fixed design samples, which later evolve based on user feedback. Designers suggested leveraging the cold-start phase to incorporate specific designs they wish to probe for user opinions. As one designer commented: ``\textit{When we set up the initial database, we sometimes have certain ideas in mind that we want users to react to}'' (D4). Allowing such customization could better align feedback collection with the designer's intent and design considerations. Furthermore, designers appreciated that the system enabled model fine-tuning based on user preferences, which enhanced their confidence in the generated results. However, some also reported uncertainty regarding how to initiate the adjustment process. One designer noted: ``\textit{If the system could provide some data-driven suggestions or guidance when we inspect the tree structure of attribute dimensions, it might help facilitate our workflow}'' (D2).

\section{DISCUSSION AND LIMITATION}

\subsection{Advancing a Comprehensive and User-Friendly Preference Collection Process}
\par In the user interface of \textit{DesignBridge}, user preferences were collected through multi-round interactions. While participants generally found the interaction mechanism intuitive and efficient, we observed that prolonged engagement could lead to information overload and choice fatigue~\cite{hick1952rate, eppler2008concept, herlocker2000explaining}. For instance, some users expressed a decreased willingness to provide detailed preference feedback after several rounds of brushing, potentially affecting the quality of the collected data. To address this, one promising direction is the adoption of dynamically adaptive interaction strategies that leverage users' prior inputs to recommend more relevant and manageable options in subsequent rounds. Such strategies can help streamline the elicitation process and support users in articulating their preferences more effectively. However, it is critical to ensure that these suggestions remain non-intrusive and avoid introducing bias that may compromise the authenticity of user expression.
\par Moreover, the system could gradually learn and adapt to individual users' expression habits and interaction behaviors, enabling a more personalized and user-sensitive interaction experience. Together, these enhancements have the potential to improve the overall preference elicitation process, supporting the collection of more accurate, meaningful, and comprehensive user preferences. In the current system, preference elicitation initially relies on a fixed set of example images intentionally selected by designers to reflect the intended scope of variation for a given design task. While this choice helps align early-stage preference probing with designer intent, a potential direction for future work is to allow designers or users to provide external reference collections, which \textit{DesignBridge} could map into the predefined design space to support a more dynamic balance between exploration and exploitation.

\subsection{Facilitating fine-grained design iteration}
\par \textit{DesignBridge} supports the iterative co-design process by enabling designers to explore and diverge from their initial design frameworks, efficiently capturing both local and global user preferences through recommendation algorithms. It then facilitates convergence by generating design outputs via a multidimensional preference integration approach. In this process, the design space functions as a semantic bridge, linking designers' creative principles with users' individualized preferences. While we provide a generalized design space model, designers are encouraged to adapt our proposed construction method to develop personalized design spaces and generative models tailored to their specific fashion design tasks.

\par Unlike prior work~\cite{arzberger2022triggered,chandrasegaran2025synthetic}, where designers engaged in dialogue-based interactions with users or chatbots to obtain design insights, \textit{DesignBridge} employs a multidimensional preference collection strategy. Consensus scores and the PPNN model capture users' local and global preferences, enabling designers to derive quantified, multidimensional insights into user needs. These preferences guide model fine-tuning and prompt formulation, whose relative influence evolves throughout the iterative design process to achieve nuanced and adaptive preference integration. Effective prompt-based generation, however, still requires highly accurate and detailed descriptions to align with the designer’s intent. As designer D2 noted, ``\textit{I sometimes wish to manually sketch the designs I have in mind to gather user feedback, which aligns more with my usual workflow.}'' This observation suggests the value of integrating sketch-based image generation~\cite{lin2025sketchflex,zhang2023sketch2realgan,wu2023styleme,wu2024stylewe} in future iterations to provide designers with more intuitive, domain-familiar creative control.

\subsection{Establishing a Sustainable Iterative User Preference Ecosystem}
\par Existing approaches~\cite{jawaheer2014modeling, markova2012analysis} often rely on users' fragmented and superficial expressions of preference, making it difficult to systematically model and retain their intent. As a result, each design iteration typically starts from scratch, without building upon a cumulative understanding of long-term user preferences.

\par To address this, our approach leverages a PPNN, which dynamically accumulates and refines a comprehensive model of user preferences over successive interactions. This mechanism transforms isolated feedback into a persistent and actionable user profile. For new design tasks targeting users with similar characteristics, this accumulated knowledge can be inherited, enabling more efficient and targeted design iterations. At the local level, we incorporate a LoRA-based model trained on individual attribute dimensions, allowing for fine-grained modeling of specific user preferences. The modular nature of LoRA allows designers to disassemble, recombine, and transfer preference modules across user groups and design tasks, enabling flexible customization, reuse, and adaptation of local preferences.

\par Looking ahead, it is promising to explore designer collaboration centered around these evolving user preference models~\cite{wu2024stylewe, bhalla2023knowledge}. By grounding collaboration in a shared understanding of user needs, designers can use the system's preference views and attribute-level analyses to bridge diverse design styles and foster richer dialogue. This shared ``preference consensus'' can not only drive more innovative design outcomes but also reposition user preferences as an active, perceivable, and sharable asset within the broader co-design ecosystem.

\subsection{Generalizability}
\label{sec:general}
\par Although our study focuses on upper-body garments within the fashion design domain, the three-stage workflow of \textit{DesignBridge} can be extended to a broader range of applications, including other clothing types such as lower-body garments and accessories, as well as related domains like interior design and furniture. Adapting the framework to these contexts would involve redefining the design space, selecting appropriate data sources, and aligning designer input formats with domain-specific requirements. Similarly, mechanisms for collecting user preferences would need to be tailored to the interaction styles and decision criteria central to each field. These adjustments maintain the integrity of the workflow while enabling its application across diverse design contexts.

\par Beyond the overall workflow, the techniques developed in this work provide a transferable foundation. The structured design space offers a systematic approach to translating complex creative concepts into well-defined attributes, which can be reformulated for other domains. The hybrid preference modeling, which integrates explicit judgments, localized feedback, and personalized recommendations, serves as a versatile template for capturing nuanced user intent. Additionally, the preference-guided fine-tuning strategy, based on modular LoRA training, enables the incorporation of user signals into generative models while preserving domain semantics. Together, these elements illustrate how \textit{DesignBridge} can be generalized beyond fashion design, supporting effective expert-user collaboration across a variety of creative practices.

\subsection{Scalability}

\par The current implementation of \textit{DesignBridge} has been evaluated in relatively small-scale co-design settings, which effectively validate user interaction and preference expression in personalized design scenarios. However, as participation scales up and design objectives shift from individual customization toward more collective or population-level design considerations, how such interactions and preference representations should be adapted remains an open question.

\par From a system design perspective, this shift implies that mechanisms originally designed to support fine-grained personalization must increasingly accommodate the aggregation, comparison, and reconciliation of heterogeneous user preferences. In this regard, the Attribute Consensus Score may be better suited to larger user populations, as it operates on aggregated preference signals at the attribute level rather than relying on direct comparisons between individual users, allowing consensus patterns to emerge as the volume of feedback increases. In contrast, the Preference Tree is more sensitive to scale. While attribute-level inspection and manual pruning support transparency in small-scale settings, these operations become increasingly difficult to manage as the number of feedback instances grows. In future iterations, we plan to explore ways to restructure the preference tree through higher-level abstraction or automated summarization, with the goal of maintaining interpretability while reducing designer workload at larger scales.

\subsection{Limitation and Future Work}

\par Our work has several limitations. First, the study was conducted in relatively small-scale co-design settings, which may limit the ecological validity of the findings in comparison to real-world scenarios involving larger user populations. Future work will examine \textit{DesignBridge} in larger-scale co-design contexts to better understand how the proposed interaction mechanisms and preference representations behave as participation increases.

\par Second, while the initial fine-tuning of the image generation model relied on a substantial amount of data, all data originated from the same e-commerce platform. As both the images and their corresponding textual descriptions exhibited a high degree of uniformity, this homogeneity may have constrained the diversity of the generated outputs. To address this, we plan to incorporate data from a broader range of sources in future iterations.
\par Third, the backend model on the user side was mainly designed to improve interaction experience and simplify preference collection. Therefore, our user study focused on collecting user ratings and qualitative feedback through questionnaires and interviews, without conducting a technical evaluation of these models’ performance.

\par Finally, the current exploration of the design space in \textit{DesignBridge} is limited to upper-body garments.
Accordingly, while the generalizability of \textit{DesignBridge} has been discussed conceptually in \autoref{sec:general}, its applicability to other fashion categories, such as lower-body clothing or footwear, has not yet been empirically validated. Notably, we employed an LLM to assist in annotating the design space using the Farfetch dataset. Although expert oversight was implemented to ensure annotation accuracy, the inherent limitations of the language model's prior knowledge required frequent human verification and correction, making the process labor-intensive. Moving forward, we aim to utilize the curated annotations to fine-tune the language model or train alternative models, enabling more efficient, potentially unsupervised annotation at scale.

\section{CONCLUSION}
\par In this study, we introduce \textit{DesignBridge}, an interactive system designed to facilitate collaboration between designers and users throughout the co-design process. Drawing on insights from a formative study, we identified current co-design practices in fashion design, along with the challenges and needs. Guided by these findings, we developed a multi-interface system for both designers and users, which together support a co-design experience. The user interface enables users to explore design options via scene-based try-on visualizations and to express preferences from both overall and specific dimensions through intuitive interactions such as brushing and commenting. The designer interface allows designers to establish an initial design framework, gather feedback across multiple dimensions, and iteratively refine generative designs by incorporating user preferences. A technical evaluation demonstrates the system's effectiveness in capturing diverse aspects of the design space and integrating user preferences into the design workflow. Additionally, a user study further validates \textit{DesignBridge}'s ability to support efficient and accurate user preference expression, assist designers in analyzing user feedback, and guide informed design generation. Consequently, this work advances co-design research by showing how human–AI collaboration can extend the co-design process itself, enabling designer expertise and user preferences to be continuously elicited, mediated, and integrated across iterative design stages.


\section{GenAI Usage Disclosure}
\par This research involved the use of generative AI tools in multiple parts of this work. During system implementation, fine-tuned generative models were used to generate intermediate design variations for eliciting user preferences, as well as final design outputs aligned with designers’ preferences. In addition, the teaser image (\autoref{fig:teaser}) was co-created with ChatGPT-4o to visualize the overall workflow, but the conceptual composition, aesthetic direction, and final content were fully determined and refined by the authors. Further, the authors used ChatGPT exclusively for linguistic polishing and grammar improvement during manuscript preparation. All generated text was thoroughly reviewed, revised, and validated by the authors, who take full responsibility for the final content.

\begin{acks}
We gratefully acknowledge the anonymous reviewers for their insightful feedback. This research was supported by the National Natural Science Foundation of China (No. 62372298), the Shanghai Engineering Research Center of Intelligent Vision and Imaging, the Shanghai Frontiers Science Center of Human-centered Artificial Intelligence (ShangHAI), and the MoE Key Laboratory of Intelligent Perception and Human-Machine Collaboration (KLIP-HuMaCo).
\end{acks}

\bibliographystyle{ACM-Reference-Format}
\bibliography{sample-base}

\appendix
\newpage
\onecolumn
\section{Interview Questions}
\label{interview-questions}
\begin{table}[h]
    \centering
    \caption{Interview questions for designers, users, and design researchers in formative study.}
    \begin{tabular}{p{4cm} p{11.5cm}}
        \toprule
        \textbf{Participant} & \textbf{Interview Questions} \\
        \midrule
        Designer & 
        \begin{itemize}
            \item Can you describe your past experiences collaborating with users in the design process?
            \item How do you typically engage users in the co-design process, and what strategies do you use?
            \item What are the main challenges in guiding users to articulate their needs and preferences?
            \item What tools, platforms, or methods have you used to support designer-user collaboration? Were they effective?
            \item How do you measure the success of co-design projects in terms of both user satisfaction and design quality?
        \end{itemize} \\
        
        User & 
        \begin{itemize}
            \item Can you describe your experience participating in a co-design process? How were you involved?
            \item How did the designer facilitate your input, and did you feel that your contributions were valued?
            \item Were there any moments when you felt it was difficult to express your needs?
            \item To what extent do you think your input was accurately reflected in the final design outcome?
            \item What challenges did you face in communicating your preferences and expectations?
        \end{itemize} \\
        
        \bottomrule
    \end{tabular}
    \label{tab:formative-study}
\end{table}

\clearpage
\section{The Details of the Garment Design Space}
\label{sec:design space}
\begin{table*}[h]
    \centering
    \caption{The details of design space, including the dimensions, attributes, and corresponding explanations.}
    \begin{tabular}{p{3cm}p{5cm}p{5cm}}
        \hline
        \textbf{Dimension} & \textbf{Attributes} & \textbf{Description} \\
        \hline
        Type & Shirt, Sportswear, Jacket, Sweater, T-shirt, Hoodie, Coat & The fundamental category of a garment. \\
        \hline
        Sleeve Length & Sleeveless, Short, Long  & The extent of arm coverage, characterized by its absence or varying lengths.\\
        \hline
        Collar Shape & Fur, Stand, V, Round, Lapel, High, Hoodie & The structural component of a garment’s neckline, encompassing its shape, opening, and height.\\
        \hline
        Wearing Style & Breasted, Pullover, Zipper & Fastening mechanism that facilitates garment wearability.\\
        \hline
        Pattern Style & Pure, Grid, Dot, Floral, Cross Stripe, Vertical Stripe, Number and Letter & The visual elements present on a garment’s surface, including prints, textures, and decorative motifs.  \\
        \hline
        Pattern Arrangement & Random, Focus, Repeat & The structural layout or distribution of a pattern.  \\
        \hline
        Material & Blending, Knit, Silk, Leather, Denim, Cotton, Chemical Fiber, Flax, Woolen & The composition and textile structure of the garment.  \\
        \hline
        Color Category & Monochromatic, Polychromatic, Dual-tone & Classification of colors based on their fundamental composition. \\
        \hline
        Specific Colors & Red, Orange, Yellow, Green, Blue, Purple, White, Black, Gray & The specific colors present within a color category. \\
        \hline
    \end{tabular}
    \label{tab:design space}
\end{table*}

\clearpage
\section{Prompt Template for Generating Virtual Try-on Background}
\label{sec:Prompt-tryon}
\begin{table}[h]
    \centering
    \caption{Prompt template for generating virtual try-on background.}
    \begin{tabular}{p{0.9\textwidth}}
        \hline
        \textbf{Prompt} \\
        \hline
        You are an expert at polishing text prompts. Based on the following task description, refine and enhance the text to make it suitable for high-quality background generation. \\
        
        \textbf{Task:} \\
        I want to generate a high quality background for virtual try-on. The background should be suitable for [Scene] scene. Please help me polish my text prompt to make it suitable for background generation. \\
        
        \textbf{Polished Prompt Example:} \\
        A vibrant and inviting spring scene. The backdrop should evoke the essence of warm spring days with elements like clear blue skies, lush greenery, and gently swaying trees. \\
        
        \textbf{Instructions:} \\
        - Preserve the intent of the original task while enhancing clarity, vividness, and visual richness. \\
        - Make the scene description evocative and suitable for AI-based background generation tools. \\
        - Keep the language natural, descriptive, and concise. \\
        
        \hline
    \end{tabular}
    \label{tab:tryon_prompt_template}
\end{table}

\newpage
\onecolumn
\section{Prompt Template for Garment Attributes Filtering}
\label{sec:Prompt-filter}
\begin{table}[h]
    \centering
    \caption{Prompt template for garment attributes filtering.}
    \begin{tabular}{p{0.9\textwidth}}
        \hline
        \textbf{Prompt} \\
        \hline
        You are an expert at designing garments and matching different scenes. Based on the following task description, refine and execute the filtering process with scene-appropriate reasoning. \\
        
        \textbf{Task:} \\
        I have a list of words describing [dimension] of a garment. I want to know which of these words are suitable to wear in a specific scene with detailed descriptions. Please help me keep [number] words remained, which are suitable for a [scene] scene, a [type] type, and a [principle] principle. \\
        
        \textbf{Sample Response:} \\
        Based on the criteria you provided for a frozen winter day scene, here are the two most suitable options: \\
        1. \textbf{Hoodie:} This option fits perfectly, especially for a frozen winter day attire. It is suitable for casual wear and comfortable in cold weather. \\
        2. \textbf{Sweater:} Considering frozen winter scene, a woolen sweater with a high collar can match the description. You can find warm sweater in various styles, including those that can be worn in cold conditions if styled appropriately. \\
        These two options should fit your specified scenario and condition. \\
        
        \textbf{Instructions:} \\
        - Evaluate each word in the given list according to the specified scene, type, and design principle. \\
        - Select and retain only the most relevant options based on contextual fit. \\
        - Provide detailed justification for each retained word. \\
        - Maintain a descriptive, professional tone with a clear focus on fashion suitability and practicality. \\
        
        \hline
    \end{tabular}
    \label{tab:garment_filter_prompt}
\end{table}

\section{Prompt Template for Avatar Generation}
\label{sec:Prompt-avatar}
\begin{table}[h]
    \centering
    \caption{Prompt template for avatar generation.}
    \renewcommand{\arraystretch}{1.5}
    \begin{tabular}{p{0.9\textwidth}}
        \hline
        \textbf{Prompt} \\
        \hline
        A man (A woman), he (she) is [height] in height (cm) and [weight] in weight (kg). He (she) is standing gracefully in the center of the frame. His (her) upper body is fully visible, wearing a short sleeve t-shirt with elegant details. \\
        
        He (she) has a gentle smile on his (her) face. The focus is on his (her) upper body, with only a slight glimpse of his (her) trousers visible at the bottom of the frame. \\
        
        \textbf{Photographic Details:} \\
        The image is taken with a Canon EOS camera, using a SIGMA Art Lens 35mm F1.4, set at ISO 200 and a shutter speed of 1/2000. The image captures every detail in stunning clarity and realism, with a high-quality, cinematic feel. \\
        
        The image is taken from the front side. His (her) body should face towards the front. It remains unobstructed in front of the body. \\
        \hline
    \end{tabular}
    \label{tab:avatar_prompt}
\end{table}

\newpage
\onecolumn
\section{Prompt template for garment generation in Stage One}
\label{sec:Prompt-stage1}
\begin{table}[h]
    \centering
    \caption{Prompt template for garment generation in Stage One.}
    \renewcommand{\arraystretch}{1.5}
    \begin{tabular}{p{0.9\textwidth}}
        \hline
        \textbf{Prompt} \\
        \hline
        An image of a clear master piece of real garment without a model, with a white background. There shouldn't a person or more than one garment in the image. \\
        
        The garment is a [Type] featuring [Sleeve Length] sleeves and a [Collar Shape] collar. \\
        
        It is worn in a [Wearing Style] style and has a [Pattern Style] pattern in a [Pattern Arrangement] arrangement. \\
        
        The garment comes in a [Color Category] color scheme, with specific colors including [Specific Colors]. The material is [Material]. \\
        \hline
    \end{tabular}
    \label{tab:lora_prompt}
\end{table}

\section{Prompt Template for Garment Generation in Stage Three}
\label{sec:Prompt-stage3}
\begin{table}[h]
    \centering
    \caption{Prompt template for garment generation in Stage Three.}
    \renewcommand{\arraystretch}{1.5}
    \begin{tabular}{p{0.9\textwidth}}
        \hline
        \textbf{Prompt} \\
        \hline
        An image of a clear master piece of real garment without a model, with a white background. There shouldn't a person or more than one garment in the image. \\
        
        The garment is a [Type] featuring [Sleeve Length] sleeves and a [Collar Shape] collar. \\
        
        It is worn in a [Wearing Style] style and has a [Pattern Style] pattern in a [Pattern Arrangement] arrangement. \\
        
        The garment comes in a [Color Category] color scheme, with specific colors including [Specific Colors]. The material is [Material]. \\

        For [Dimension] part, with detailed descriptions of [Detailed Prompt]. \\
        \hline
    \end{tabular}
    \label{tab:garment_generation_prompt_after_first}
\end{table}

\newpage

\section{Detailed Information of Participants in the User Study}
\label{sec:participants-userstudy}
\begin{table*}[h]
\centering
\footnotesize
\setlength{\tabcolsep}{4pt}
\renewcommand{\arraystretch}{1.35}
\caption{Detailed participant information in the user study.}
\label{tab:participants}

\begin{tabular*}{\textwidth}{@{\extracolsep{\fill}} c c p{3.0cm} c c c c p{3.0cm} @{}}
\toprule
\textbf{Condition} & \textbf{Group} & \textbf{Design Theme} & \textbf{ID} & \textbf{Role} & \textbf{Gender} & \textbf{Age} & \textbf{Experience / Familiarity} \\
\midrule
\multirow{21}{*}{DesignBridge} & \multirow{4}{*}{G1} & \multirow{4}{*}{Minimalist Casual Wear} & D1 & Designer & F & 34 & 7 years \\
 &  &  & U1 & User & M & 23 & Moderate \\
 &  &  & U2 & User & M & 25 & High \\
 &  &  & U3 & User & M & 24 & Moderate \\
\cline{2-8}
 & \multirow{4}{*}{G2} & \multirow{4}{*}{Modern Office Attire} & D2 & Designer & M & 31 & 5 years \\
 &  &  & U4 & User & F & 26 & Moderate \\
 &  &  & U5 & User & F & 27 & High \\
 &  &  & U6 & User & F & 26 & Moderate \\
\cline{2-8}
 & \multirow{5}{*}{G3} & \multirow{5}{*}{Resort Travel Wear} & D3 & Designer & F & 32 & 4 years \\
 &  &  & U7 & User & F & 26 & Moderate \\
 &  &  & U8 & User & F & 25 & High \\
 &  &  & U9 & User & F & 24 & Moderate \\
 &  &  & U10 & User & F & 27 & High \\
\cline{2-8}
 & \multirow{4}{*}{G4} & \multirow{4}{*}{Sporty Athleisure} & D4 & Designer & M & 28 & 4 years \\
 &  &  & U11 & User & M & 23 & Moderate \\
 &  &  & U12 & User & M & 25 & High \\
 &  &  & U13 & User & M & 24 & Moderate \\
\cline{2-8}
 & \multirow{4}{*}{G5} & \multirow{4}{*}{Gender-neutral Everyday Wear} & D5 & Designer & F & 33 & 6 years \\
 &  &  & U14 & User & F & 25 & Moderate \\
 &  &  & U15 & User & M & 27 & High \\
 &  &  & U16 & User & M & 26 & Moderate \\
\hline
\multirow{21}{*}{Baseline} & \multirow{4}{*}{G6} & \multirow{4}{*}{Minimalist Casual Wear} & D6 & Designer & F & 33 & 5 years \\
 &  &  & U17 & User & F & 23 & Moderate \\
 &  &  & U18 & User & F & 25 & High \\
 &  &  & U19 & User & F & 26 & Moderate \\
\cline{2-8}
 & \multirow{4}{*}{G7} & \multirow{4}{*}{Modern Office Attire} & D7 & Designer & F & 32 & 6 years \\
 &  &  & U20 & User & F & 27 & Moderate \\
 &  &  & U21 & User & F & 28 & Moderate \\
 &  &  & U22 & User & F & 31 & High \\
\cline{2-8}
 & \multirow{4}{*}{G8} & \multirow{4}{*}{Resort Travel Wear} & D8 & Designer & M & 31 & 4 years \\
 &  &  & U23 & User & M & 27 & High \\
 &  &  & U24 & User & M & 25 & Moderate \\
 &  &  & U25 & User & M & 26 & High \\
\cline{2-8}
 & \multirow{4}{*}{G9} & \multirow{4}{*}{Sporty Athleisure} & D9 & Designer & F & 31 & 5 years \\
 &  &  & U26 & User & M & 22 & Moderate \\
 &  &  & U27 & User & M & 25 & High \\
 &  &  & U28 & User & M & 24 & Moderate \\
\cline{2-8}
 & \multirow{5}{*}{G10} & \multirow{5}{*}{Gender-neutral Everyday Wear} & D10 & Designer & M & 33 & 5 years \\
 &  &  & U29 & User & M & 23 & Moderate \\
 &  &  & U30 & User & F & 25 & Moderate \\
 &  &  & U31 & User & M & 26 & High \\
 &  &  & U32 & User & F & 27 & High \\
\bottomrule
\end{tabular*}

\vspace{3pt}
\end{table*}

\end{document}